\documentclass[aps,pra,article,onecolumn,showpacs,preprintnumbers,amsmath,amssymb,superscriptaddress]{revtex4-1}
\date{\today}
\usepackage{epsfig}
\usepackage{subfigure}
\usepackage{graphicx}
\usepackage{dcolumn}
\usepackage{bm}
\usepackage[colorlinks,linkcolor=blue,hyperindex,CJKbookmarks]{hyperref}
\usepackage{float}
\usepackage{hyperref}
\usepackage{comment}
\usepackage{color}
\usepackage{psfrag}
\usepackage{color}
\usepackage{amsmath}
\usepackage{amssymb}
\usepackage{enumerate}

\begin{document}

\title{Propagation of an orbiton in the antiferromagnets:\\theory and experimental verification}

\author{Krzysztof Wohlfeld}
\affiliation{Institute of Theoretical Physics, Faculty of Physics, University of Warsaw, Pasteura 5, PL-02093 Warsaw, Poland}

\date{\today}
\begin{abstract}
In this short review, which is based on the works published between 2011 and 2016, 
we discuss the problem of the propagation of a collective orbital excitation (orbiton) created in the Mott insulating and
antiferromagnetic ground state. On the theoretical side, the problem is solved by mapping a Kugel-Khomskii spin-orbital model
describing an orbiton moving in an antiferromagnet onto an effective $t$--$J$ model with a `single hole' moving
in an antiferromagnet. The most important consequence of the existence of the above mapping is the fractionalisation
of the electron's spin and orbital degree of freedom in the 1D antiferromagnets---a spin-orbital separation phenomenon that is similar
to the spin-charge separation in 1D but corresponds to an exotic regime where spinons are faster than holons. 
Besides a detailed explanation and benchmarking of the mapping, in this review we also discuss its application to several
relatively realistic spin-orbital models, that are able to describe the experimentally
observed orbital excitation spectra of copper and iridium oxides. 
\end{abstract}
\pacs{xxxx}
\maketitle

\section{Preface: Beyond the Noninteracting Collective Excitations}

`More is different' is a celebrated idea which was postulated by Philip Anderson in the early 1970s~\cite{Anderson1972}. It summarises
the notorious feature of condensed matter physics: when one puts together a macroscopic number of interacting
particles (e.g. electrons, atoms, or molecules), then the resulting state of matter formed by these particles has completely different
properties than its constituents. Probably the most famous implementation of that idea is the concept of 
{\it order and collective excitations}. It turns out that the ground states of a number of condensed matter systems
show order while its low-lying excited states can be most easily understood in terms of noninteracting quasiparticles,
the so-called collective excitations~\cite{Khomskii2010}. A well-known example of that concept is the `standard' crystal.
In this solid material its constituents form a periodic arrangement (crystal lattice) leading to an ordered ground state. 
At the same time the low-lying excited states of the crystal can be described in terms of collective vibrations of all atoms in a crystal---the phonons. 

More exotic realisations of the concept of 
`order and collective excitations'  can be found in 
crystals with local and anomalously strong electron-electron interactions.
Such strongly correlated electron systems are ubiquitous to 
several transition metal compounds (typically oxides or fluorides) with deep lattice potentials~\cite{Imada1998}.
Although these compounds may exhibit very distinct characteristics, 
what they all have in common is that the well-known properties of a single electron,
such as for instance the electric charge or the spin angular momentum, may need to be considered as 
forming separate `entities'~\cite{Khomskii2010, Auerbach1994}. While this is another facet of the idea that `more is different', this also 
means that the collective excitations supported by these systems 
may for instance solely carry the electron's spin angular momentum---but no charge.

A nice example of such physics can be found in one of the most-studied class of transition metal compounds---the undoped quasi-2D copper oxides
(La$_2$CuO$_4$, CaCuO$_2$, etc.).
Here the electron-electron interactions, as for instance modelled by the 2D square-lattice Hubbard Hamiltonian~\cite{Lee2006},
are so strong that they tend to localize electrons on the copper ions leading to the Mott insulating ground state of such a 
system\footnote{One should note that the Mott insulators can be split into two separate classes~\cite{Zaanen1985}: (i) Mott-Hubbard insulators, (ii) charge transfer insulators.
The copper oxides that are studied in this review are charge transfer insulators~\cite{Imada1998}.}.
Nevertheless, still virtual electronic motions are allowed, leading to the spin exchange processes: 
the effective interaction solely between the spin degrees of freedom of the electron~\cite{Khomskii2010, Auerbach1994}. The latter one can be modelled
by a spin-only Heisenberg-like Hamiltonian~\cite{Imada1998, Coldea2001}. The (zero temperature) ground state of 
that model shows antiferromagnetic (AF) long range order and the low lying excited states can be well-described in terms 
of noninteracting bosonic quasiparticles (called magnons)~\cite{Auerbach1994}.
It turns out that the theoretically calculated dispersion of these collective magnetic excitations very well agrees with the one that was 
measured experimentally~\cite{Coldea2001, Braicovich2009a, Peng2017}.

The fact that the low lying excitations of the undoped copper oxides can 
be successfully described by an essentially noninteracting theory is very appealing.
Unfortunately, in the field of correlated electron systems 
such a situation is not that common.
This is for instance the case of the so-called `high-Tc problem', i.e. the question of the origin
of the high-temperature superconductivity observed upon hole- or electron-doping the (already-mentioned above) copper oxides~\cite{Imada1998, Lee2006}.
Here the concept of `order and collective excitations' seems to break down
and, despite more than 30 years of intensive research, the nature of the ground state
of this strongly interacting system is not understood. 
Using the words of J.~Zaanen~\cite{Zaanen2013} one can say that such a ground state
``(...) is some form of highly
collective `quantum soup' exhibiting simple
yet mysterious properties such as the `local
quantum criticality' (...)''.

In this review, we intend to understand a problem that lies in between these two extreme cases. That is, on one hand, far simpler than understanding the nature 
of that highly collective `quantum soup'.
On the other, it is significantly more complex than a simple noninteracting theory of collective excitations.
As will be shown below, on the way to understand the main problem posed in this review, which concerns the 
nature of the collective orbital excitations (also called orbitons, see Sec.~II below) that could be observed in the transition metal compounds,
we will need to tackle the problem of {\it a condensed matter system which supports collective excitations of different kinds
that are interacting with each other}. 

The review is organised as follows. Sec.~II introduces the basic concepts and problems of the so-called `orbital physics' which
are needed to define the main problem that is discussed in this review---i.e. the propagation of an orbiton in an antiferromagnet (Sec.~III). 
Next, we discuss the solution to that problem as obtained for a `minimal' spin-orbital model (Secs.~IV-VI).
Sections~VII-IX show the solutions to that problem as found for more realistic models, which can describe
the physics observed in several transition metal oxides. Finally, in Sec.~X we summarize the findings and give a short outlook at the possible future studies.

\section{Introduction: What Are Orbitons and Where to Find Them?}

Quite often it is only the electron's {\it spin} angular momentum that needs to be taken into account in the electronic description of a solid.
The {\it orbital} angular momentum, associated with binding of an electron to the nucleus of one of the 
atoms forming the solid, is often neglected. This is typically the case of weak lattice potentials and some `conventional' metals or band insulators.
However, in the case of deep lattice potentials found in the transition metal compounds 
the electrons `live' in the atomic-like wavefunctions carrying distinct orbital quantum numbers. 
To put it differently: the noninteracting Hamiltonian of these solids can be written in terms of the atomic-like wavefunctions that
are eigenstates of the orbital angular momentum operator~\cite{Griffith1964, Kugel1982, Tokura2000}.
How does the nonzero orbital angular momentum of an electron may affect the many-body (correlated) 
physics present in these compounds? Let us illustrate that on a simple, and probably the most famous, example---KCuF$_3$:

At first sight this copper fluoride with a perovskite structure seems to be somewhat similar to the mentioned in the Preface quasi-2D copper oxides.
In the ionic picture the nominal valence of the copper ion is $2+$ in this compound which
leads to the the commensurate filling of one hole in the $3d$ shell of the copper ion. This, together with the strong on-site electron-electron repulsion,
might in principle lead to a very similar physics as the one found in the quasi-2D cuprates and already described above. One could think that the 3D version of the 
Hubbard model should be good enough in describing the low energy electronic properties of this solid. This could then lead to the charge localization in the Mott insulating state 
and an effective (3D) spin-only Heisenberg model with antiferromagnetic long range order and well-defined magnon excitations in low temperatures. 

However, it turns out that in reality the physics of KCuF$_3$ is quite different. The culprit is the orbital degree of freedom or, more precisely, the so-called `orbital degeneracy'. 
Unlike the quasi-2D cuprates which basically show tetragonal symmetry, the copper ions in KCuF$_3$ can be regarded as being in a cubic cage formed by the
six nearest neighbor oxygen ions. Since the wave functions pointing toward fluorines have higher energy in comparison with those
pointing between them, two of the lowest lying (in hole language) $3d$ orbitals are degenerate in this picture. These are the cubic harmonics
$d_{x^2-y^2}$ and $d_{3z^2-r^2}$ (jointly called as the `$e_g$ orbitals'). Such degeneracy strongly modifies the many-body physics:
while the charge is still localized as a result of strong electron-electron repulsion and commensurability, 
the virtual electronic motions are now permissible between electrons with different spin {\it and} orbital quantum number.
Consequently, instead of a simple spin-only Heisenberg model, the effective low energy model  (often called spin-orbital or Kugel-Khomskii) contains
both spin and orbital degrees of freedom model Hamiltonian~\cite{Kugel1982}] and
takes the following form
\begin{align}\label{eq:KK}
\mathcal{H}= J \sum_{\langle {\bf i}, {\bf  j} \rangle || \Gamma} \left({\bf S}_{\bf i} \cdot {\bf S}_{\bf j} + A \right) \left(B\ {T}^z_{\bf i}  {T}^z_{\bf j} 
+ C\ {T}^x_{\bf i}  {T}^x_{\bf j} 
+D_{\alpha}\ T^{\alpha}_{\bf i} + E \right),
\end{align}
where $J$ is the spin-orbital exchange, $A, B, C, D_{\alpha}, E$ are constants of the model (which depend on the orbital hoppings and on the ratio
of the Hund's exchange to the on-site Coulomb repulsion), and $\Gamma=a, b, c$ is a particular direction in the cubic lattice formed by copper ions,
Crucially, ${\bf T}_{\bf i}$ is a $T=1/2$ orbital (pseudo)spin operator on site ${\bf i}$ which is a three component vector ($T^\alpha_{\bf i} $ where $\alpha=x, y, z$)
and which takes care of the orbital degrees of freedom (the two eigenvalues of $ T^z_i$  correspond to one of the two $e_g$ orbitals being occupied),
cf.~\cite{Kugel1982, Oles2000} for further details.
Note that the interaction between orbital degrees of freedom lacks the $SU(2)$ symmetry which is a consequence of the finite Hund's exchange and the distinct hopping 
elements between different orbitals. This situation is typical to several spin-orbital models~\cite{Oles2005}.

Finding the ground state of the inherently interacting model (\ref{eq:KK}) is a complex task. It is still a matter of active research~\cite{Brzezicki2011, Brzezicki2013, Czarnik2017}, despite the 35 years that has passed since the seminal paper by Kugel and Khomskii~\cite{Kugel1982}. Nevertheless, a relatively good insight into the problem can be obtained by firstly considering just one plane of this 3D model (which corresponds to the $ab$ plane of KCuF$_3$) and by going to the orbital basis formed by the $d_{x^2-z^2}$ and the $d_{y^2-z^2}$ orbitals, i.e. without off-diagonal hoppings between orbitals in the $ab$ plane.
Then, a simple of the spin-orbital model (\ref{eq:KK}) suggests that the ground state of that Hamiltonian should show alternating-orbital (AO) order in the $ab$ plane [cf.~Fig.~1(a)] and a concomitant spin ferromagnetic (FM) order\footnote{While understanding in detail the origin of that type of spin and orbital order in this model is somewhat intricate,
it is important to stress that such a tendency to the AO and FM order is triggered by the finite Hund's exchange that is inherently present in spin-orbital model Eq.~(\ref{eq:KK}), cf.~Refs.~\cite{Kugel1982, Feiner1997} and in particular~Fig.~12 of Ref.~\cite{Kugel1982}.}. On the other hand, a similar analysis for the $c$ direction yields the same energy for the ferroorbital (FO) order and AO order, both accompanied
by the AF order -- hence two different types of orbital ordering are realized in KCuF$_3$, cf.~Fig.~1(a). We should note that, although it is still not completely clear what the ground state of model (\ref{eq:KK}) is and whether the Jahn-Teller effect may also play a significant
role in stabilizing the observed orbital oder in KCuF$_3$ below $T\sim800K$~\cite{Kadota1967, Pavarini2008, Lee2011}, it is generally agreed that for the realistic values of the parameters the ground state of such a spin-orbital model supports long-range spin and orbital order~\cite{Oles2000, WenLong2007, Brzezicki2013}.

\begin{figure}[t!]
\includegraphics[width=0.99\columnwidth]{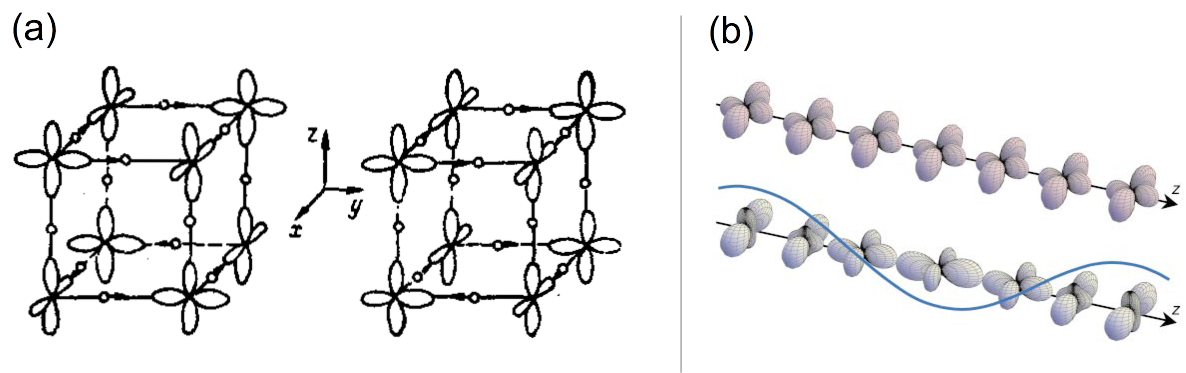}
\caption{{\bf Artist's view of the orbital order and of the orbiton}
(a) Artist's view of the orbital order in KCuF$_3$ type
as predicted by the the spin-orbital exchange model Eq.~(\ref{eq:KK}) with realistic parameters [AO order formed by the $d_{x^2-z^2}$ and the $d_{y^2-z^2}$ orbitals 
in the plane and either FO or AO order along $c$ direction]; Left and right panels show
two equivalent types of ordering (interestingly, both are observed in KCuF$_3$).
(b) Artist's view of the propagation of the collective orbital excitation in an FO system;
Top panel shows the  FO ground state; Bottom panel show the orbital state with one orbiton at a particular momentum. 
[Panel (a) and caption adopted from Ref.~\cite{Kugel1982}. Panel (b) and caption adopted from Ref.~\cite{Saitoh2001}]
}
\label{fig:1}
\end{figure}

As mentioned in the Preface, typically the onset of some type of order in the ground state of a condensed matter system leads to collective excitations from that ground state. 
Therefore, it should not come as a surprise that, as a result of the spin and orbital order present in the ground state of model (\ref{eq:KK}), we should expect that collective spin and
orbital excitations should appear here. Indeed, a mean-field decoupling of spins and orbitals in Eq.~(\ref{eq:KK}) predicts that those low energy excited states which carry solely the spin
degree of freedom can indeed be described in terms of magnons~\cite{Oles2000}. This finding was confirmed by the inelastic neutron scattering experiment on KCuF$_3$~\cite{Lake2000}. The fundamental question relates to the existence of the collective orbital interactions~\cite{Kugel1973} -- the so-called orbitons (this term was probably used for the first time in~\cite{Pen1997}), cf. Fig.~1(b).
Indeed the mean-field analysis presented in~\cite{Oles2000} suggests that not only the orbitons but even joint spin-orbital collective excitations should be present in the model. 

Unfortunately, so far the orbitons have not been observed experimentally in KCuF$_3$. What could be the reasons for that? 
It seems to us that the main problem lies on the experimental side. For years there had not been a reliable experimental probe to measure orbiton dispersion~\cite{Ishihara2000, Forte2008, Marra2012}. 
Even though neutrons couple to the orbital excitations~\cite{Ishihara2004}, this cannot be easily realized 
experimentally~\cite{Shamoto2005} due to generally low energy transfers in the inelastic neutron scattering
experiments. On the other hand, Raman scattering cannot transfer momentum to orbital excitations leading
to controversies concerning the interpretation of the experimental spectra of LaMnO$_3$ -- a `sister'-compound to KCuF$_3$~\cite{Saitoh2001, Grueninger2002, Ishihara2004}. 
Finally, while the resonant inelastic x-ray scattering (RIXS)~\cite{Ament2011} may seem to be the best probe of the orbitons~\cite{Ishihara2000, Forte2008},
so far it could not detect an orbiton\footnote{The experimental RIXS spectra taken at the Cu $L$ edge of KCuF$_3$ do not show
any signatures of orbitons (unpublished: C. Mazzoli, G. Ghiringhelli, K. Wohlfeld, and T. Schmitt).} in the measured spectra of KCuF$_3$. 
The reason probably lies in the limiting energy resolution of RIXS: in order for the orbital excitation to be understood as having a collective nature (and thus be called an orbiton)  
we need to detect a finite orbiton dispersion. Since the magnitude of the orbiton dispersion is dictated by the size of the effective orbital exchange interaction, 
it usually is of the order of 10s of meV in the transition metal compounds -- e.g. simple calculation of the orbiton bandwidth estimates it to be of the order
of 10-20 meV in KCuF$_3$. Therefore, using the current RIXS resolution (ca. 55 meV at Cu $L$ edge, cf. Ref.~\cite{Peng2017})
the orbiton cannot be unambigously detected in KCuF$_3$.
 
An even more disturbing fact is that basically till 2012 (cf. Ref.~\cite{Schlappa2012} and below) the orbitons had not been unambiguously detected in any transition metal compound. First of all, as already mentioned above, the Raman scattering has not managed to detect orbitons in LaMnO$_3$, probably the most famous transition metal 
oxide with orbitally degenerate orbitals, as the earlier results~\cite{Saitoh2001} were rejected  by a consecutive study~\cite{Grueninger2002}. 
While few other studies have shown the orbiton existence in an indirect way (e.g. through the so-called Davydov splittings~\cite{Macfarlane1971} in Cr$_2$O$_3$, 
or in a pump-probe experiment in the doped manganite~\cite{Polli2007}), perhaps it was only in in the late 2000s that finally a small 
orbiton dispersion was indeed detected by RIXS on the titanates~\cite{Ulrich2009} -- however, that signature was very weak as the dispersion was basically on 
the edge of the RIXS experimental resolution. 

In this review, which is centered around the works published in the years 2011-2016, we suggest that the way to overcome the problem of finding the orbiton 
is to look at different systems: the transition metal oxides {\it without} orbital degeneracy in the ground state, i.e with relatively large crystal field splitting between orbital levels in the ionic 
picture and, therefore, having (by definition) simple FO ground states (accompanied by the AF spin state due to the Goodenough-Kanamori rules~\cite{Kanamori1959, Goodenough1963}). 
Naively, this might be a rather counterintuitive thing to do, since usually in this case one does not at all discuss the orbital degrees of freedom.
However, one should bear in mind that this does not mean that the orbital degrees of freedom are simply irrelevant: it is just that the ground state
description can be obtained without taking them into account. At the same time, the orbitals are important for the excited states in such compounds
-- and those can be accessed in an experiment (e.g. in RIXS). Crucially, the main advantage of choosing the FO-AF class of systems is as follows:
as the exchange interaction is typically much larger in the FO and AF systems than in the AO and FM systems~\footnote{This roughly follows from the following reasoning:
(i) in contrast to the AO-FM systems which usually occur for orbitally degenerate ground states in the ionic picture, 
the FO-AF are systems usually stabilized in the presence of large crystal fields, (ii) such crystal fields are to a large extent due to strong hybridization between transition metal ion and its ligands, (iii) strong hybridization means large hopping elements and therefore (potentially) strong spin and orbital exchange interactions.}, there are higher chances of detecting the orbiton dispersion using the resolution of the currently available experimental techniques.

\section{Subject of the Review: Propagation of an Orbiton in an Antiferromagnet}

We have already stressed that the search for collective orbital excitations had turned out to be unsuccessful for several years. 
As already discussed, this might be due to the small spin-orbital exchange in systems with AO order. One of the ways to overcome this 
problem is to look for systems with the orbital FO and, as a result of the Goodenough-Kanamori rules~\cite{Kanamori1959, Goodenough1963}, spin AF ground states~\footnote{One should stress that for the hypercubic lattices (that
of interest here) the ground state of the $S=\frac12$ AF Heisenberg spin subsystems can be classified as {\it ordered}:
it is spontaneously broken and long-range-ordered on a 2D or 3D square lattice~\cite{Auerbach1994} and is `algebraically' ordered~\cite{Auerbach1994, Mila2000} in one dimension (1D; due to the slowly decreasing power-law spin-spin correlations). Naturally, using such a terminology does not mean that we will {\it not} consider below the important differences between the 1D and 2D or 3D cases.}. 
However, this requires 
understanding how such a collective orbital excitation can move in the AF state -- which is the 
main subject of this review. We note already here that, due to the inherent entanglement of the spin and orbital degrees of freedom~\cite{Oles2006}
in a wide class of spin-orbital models, this turns out to be a relatively complex problem. This statement is further supported by several works dedicated to the related problems~\cite{Khaliullin1997, Khaliullin2000, Kikoin2003, Wohlfeld2009, Herzog2011, Wen-Long2012} which suggest that the spin and orbital degrees of freedom cannot be decoupled on a mean-field level in a number of spin-orbital models. While in Sec.~IV immediately below we formulate and solve this question in the context of a `minimal' spin-orbital model, 
we will discuss various aspects of this problem throughout the whole review.

\section{`Minimal' Spin-Orbital Model: Mapping onto a Single Hole Problem}

We start by postulating the Hamiltonian of such a `minimal' spin-orbital model. To this end, we require that:
(i) to simplify the matters we concentrate on the lowest number of spin and orbital degrees of freedom per site that, however, still leads to a nontrivial solution;
(ii) for the sake of the understanding we try to keep the model as close to the well-known Heisenberg model as possible; 
(iii) it has an AF ground state. This leads us to the following spin-orbital Hamiltonian with the Heisenberg [$SU(2)$--symmetric] interactions
between both spin and orbital (pseudospin) degrees of freedom:
\begin{align}\label{eq:h}
{\mathcal H} = 
4J \sum_{\langle i,j\rangle} \Bigl( {\bf S}_{i}\cdot{\bf S}_{j}+\tfrac{1}{4} \Bigr) \Bigl( {\bf T}_{i}\cdot{\bf T}_{j}+\tfrac{1}{4} \Bigr)  +  E_z \sum_{i}  T^z_i,
\end{align}
where ${\bf S}$ (${\bf T}$) are the spin (pseudospin) operators that fulfill the $SU(2)$ algebra for 
$S=1/2$ ($T=1/2$) spins (pseudospins) and $i$ and $j$ are lattice sites and on each bond 
$\langle i,\!j \rangle $. Note that $T^z_i = (n_{i a} - n_{i b})$ where the operatore $n_{i \alpha}$ counts the number of electrons in orbital
$\alpha = a, b$. The constant $J>0$ gives the energy 
scale of the spin-orbital exchange (typically, but somewhat confusingly, called superexchange) 
and $E_z$  the symmetry breaking field in the orbital sector. We note that the geometry and dimensions of the lattice needs not be defined at this stage.
\begin{figure}[t!]
\begin{center}
\includegraphics[width=0.85\columnwidth]{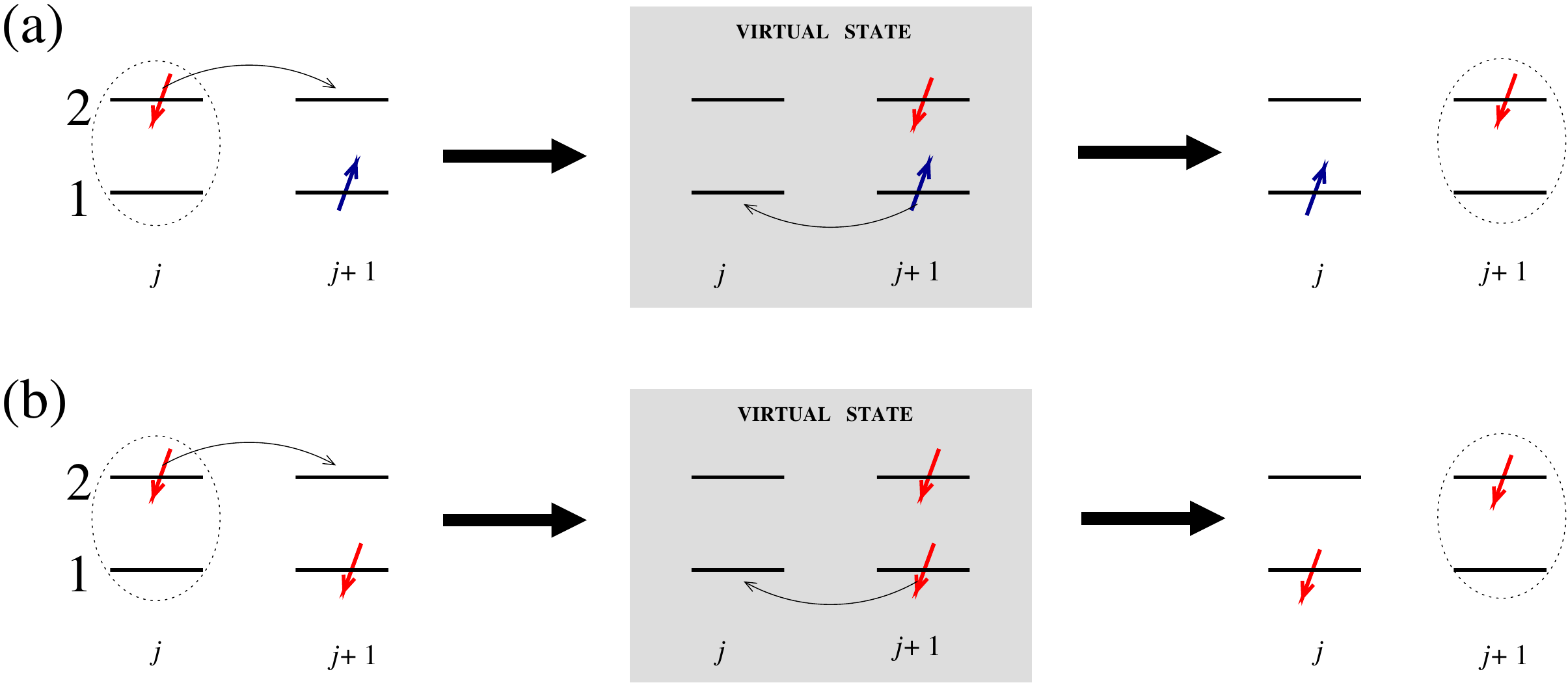}
\end{center}
\caption{{\bf Conservation of the spin on the excited orbital in the exchange process}
Two spin-orbital exchange processes moving an electron in an excited orbital 
(indicated by oval) from site $j$ to its neighbor $j+1$:
(a) and (b) describe exchange process when spins along the bond are
antiparallel or parallel, respectively, see text.
The states in the grey middle panels are not
part of the low-energy Hilbert space corresponding to
(\ref{eq:h}). These virtual excitations within the full two-orbital
\emph{Hubbard} model illustrate the origin of 
those spin-orbital exchange interactions
of Hamiltonian (\ref{eq:h}) that propagate the orbiton.
Note that the spin of the excited electron is
conserved, which is a crucial feature that enables the mapping of the
orbital problem onto the effective $t$--$J$ problem.
[Figure and caption adopted from Fig.~1 of  Ref.~\cite{Wohlfeld2011}.]
}
\label{fig:2}
\end{figure}

It is relatively easy to verify that the above Hamiltonian fulfils the first two of `our' requirements. As for the last one, which concerns the AF ground state, 
one needs to look a bit closer at the problem, for the ground state of the model obviously depends on the ratio of the model parameters $E_z/J$. In particular, once 
$E_z=0$ the spin-orbital Hamiltonian has an $SU(4)$ symmetry, even higher than the
combined $SU(2)\times SU(2)$ symmetries, which for instance in 1D results in the ground state
given by the Bethe Ansatz and composite spin-orbital gapless excitations 
in addition to the separate spin and orbital ones (see, e.g., Ref.~\cite{Li1999}). Therefore, the case $E_z= 0 $ is not of interest here (see, however, Sec. V. below). 
On the other hand, once the crystal field is much larger than the spin-orbital exchange, $E_z \gg J$, then
the ground state $| \psi \rangle = | \psi_S \rangle\otimes| \psi_O \rangle$ of Eq.~(\ref{eq:h}) {\it decouples} into the `spin' ($ | \psi_S \rangle $) and `orbital' ($ | \psi_O \rangle $) sectors.
This is because in this limit, in the `orbital sector' the ground state of the above Hamiltonian is ferroorbital (FO): $|\psi_O \rangle = |\textrm{FO}\rangle$ is a simple product state of single-site eigenstates of the $T^z_i$ operator with the $-1/2$ eigenvalue on each site $i$, while all other orbital configurations are separated from this state by a gap $\propto E_z$. At the same time, the `spin sector' is then described as an eigenstate of the spin-only  Heisenberg Hamiltonian with an effective spin exchange interaction equal to $2J$, i.e. it is
an AF Heisenberg system formed by spins in the lower-energy orbital ($| \psi_S \rangle =|\textrm{AF}\rangle$). While the precise form of the spin sector ground state depends on the 
choice of the lattice, for the most studied case of the hypercubic lattice and the $S=1/2$ case it can basically be classified as `ordered': 
it is spontaneously broken and long-range-ordered on a 2D or 3D square lattice~\cite{Auerbach1994} and is `algebraically' ordered~\cite{Mila2000} in 1D (due to the 
slowly decreasing power-law spin-spin correlations). 
We note in passing that in this case the low energy excited states can typically be described in terms of noninteracting quasiparticles (spinons in 1D  or magnons on 2D or 3D hypercubic lattices): 
the interactions between spinons (or magnons) are negligible provided their number in the AF is of the order O(1) and the system is infinite
~\cite{Karbach2000, Karbach2002}. 

Having understood that spin-orbital model (\ref{eq:h}) indeed has an AF ground state once $E_z \gg J$, we are now ready to formulate the above-mentioned 
main subject of this review in the context of this model.
To this end, we define the following spectral function which describes the dynamics of the orbital excitation added to the AF ground state as
\begin{align} \label{eq:spectral}
O(k, \omega)=\frac{1}{\pi} \lim_{\eta \rightarrow 0} 
\Im \left\langle \psi \left| T^-_{k} 
\frac{1}{\omega + E_{\psi}  - \mathcal{{H}} -
i \eta } T^+_k \right| \psi \right\rangle,
\end{align}
where $E_{\psi} $ is the energy of the ground state $ | \psi_S \rangle $. The momentum-dependent orbital excitation is given by $T^+_k=\sum_j \exp (i  k j) T^\dag_j$ [$T^-_k=(T^+_k)^\dag$] with $T^+_j$ being the orbital raising operator which promotes an electron at site $j$ from the occupied lower orbital to the empty 
higher one at the same site. Below we intend to calculate the orbiton spectral function and understand its properties.
While that in principle can be done numerically (at least in the 1D case, see Sec.~V below), here we follow a different approach.
We perform an {\it exact mapping} of the orbital problem in question onto a well-known problem of a single hole in the AF ground state. 

The crucial feature of the orbital problem, which permits such an exact mapping between these two problems, 
is related to the conservation of the spin on the excited orbital during its propagation, see Fig.~\ref{fig:2}(b). This peculiar conservation follows from the fact
that irrespectively of the spin arrangement of the two nearest neighbor spins, the spin of the electron in the excited orbital does not change after the spin-orbital
exchange process encoded in Hamiltonian (\ref{eq:h}). Consequently, the spin of the electron in the excited
orbital can be understood as being a `silent' degree of freedom and the electron in the excited orbital can be
treated as a `hole' in the AF ground state --  with the sole difference that it propagates via a spin-orbital
exchange $\propto J$ and not via the tight-binding hopping element $\propto t$. 

Mathematically, the mapping proceeds as follows:
(i) using the Jordan-Wigner transformation~\cite{Jordan1928} we replace the spins (pseudospins) with spinons (pseudospinons), respectively;
(ii) we observe that the pseudospinon-pseudospinon terms vanish (due to the fact that there is just one pseudospinon in the whole bulk);
(iii) we observe that (as a result of the above-mentioned conservation of the spin quantum number during the orbital propagation) 
one can introduce the constraint which requires that there can never be a pseudospinon and a spinon on the same site;
(iv) we express all terms containing the pseudospinons in terms of the constrained fermions (the latter ones are fermions subject
to the constraint of no double occupancies, cf. ~the $t$--$J$ model problem~\cite{Chao1977, Spalek2007}), cf. Ref.~\cite{Barnes2002};
(v) we express all remaining terms containing spinons in terms of the $S=1/2$ spins. 
It is interesting that the above procedure does not depend on the dimensionality of the problem (despite using the Jordan-Wigner
transformation, which usually is better suited for the 1D problems). On the other hand, the mapping would no longer be
exact if a spin-orbital Hamiltonian allowed for processes which do not conserve the spin quantum number during the orbital propagation
(cf. Sec.~VII below).

Altogether, we arrive at the following conclusion: It turns out that one can exactly map the problem of a single orbital excitation 
in the AF and FO ground state with its dynamics governed by the spin-orbital 
Hamiltonian [Eqs.(\ref{eq:h}-\ref{eq:spectral})] onto a problem a single hole in the AF with the dynamics governed by the appropriate $t$--$J$ model:
\begin{align} \label{eq:spectraltJ}
A(k, \omega)\!=\!\frac{1}{\pi} \lim_{\eta \rightarrow 0} 
\Im \left\langle {\tilde{\psi}}  \left| {\tilde{p}}^{\dag}_{k \uparrow}   \frac{1}{\omega \!+\!
E_{{\tilde{\psi}} }  \!-\! \mathcal{{\tilde{H}}} \!-\! E_z \!-\! i \eta } {\tilde{p}}_{k
\uparrow}  \right| {\tilde{\psi}} \right\rangle.
\end{align}
Here $ |{\tilde{\psi}} \rangle$ ($E_{{\tilde{\psi}}}$) is the ground state (energy of the ground state) of the undoped $t$--$J$ model with the 
Hamiltonian given by
\begin{align} \label{eq:tJ}
{\mathcal {\tilde{H}}}\! = &\! -\! t \!\sum_{\langle i, j \rangle, \sigma} \left( \tilde{p}^\dag_{i \sigma}
\tilde{p}_{j \sigma}\! +\! h.c. \right) 
\! +\! 2J \sum_{\langle i, j \rangle } \left( {\bf S}_{i } \cdot {\bf S}_{j} \! +\! \frac14 \tilde{n}_{i}
\tilde{n}_{j} \right), 
\end{align}
where $\tilde{p}_{j \sigma}$ act in the restricted Hilbert space without double occupancies, 
$\tilde{n}_{j } =\sum_{\sigma} \tilde{n}_{jp\sigma}$, and the hopping parameter $t$ is 
defined as $t=J$.

\begin{figure}[t!]
\begin{center}
\includegraphics[width=0.85\columnwidth]{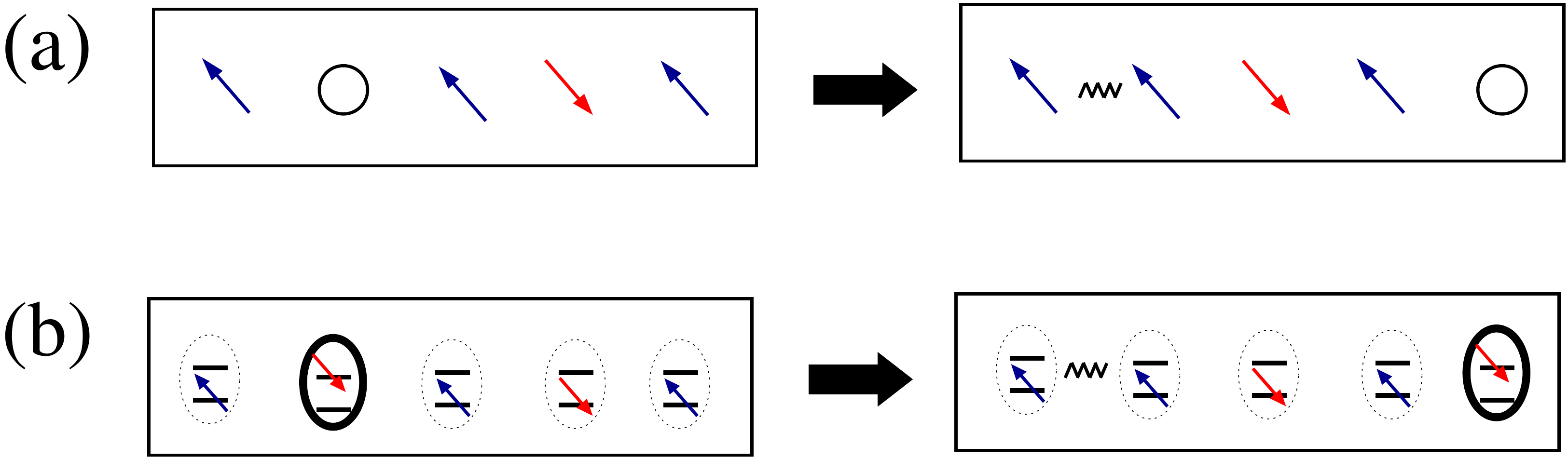}
\end{center}
\caption{{\bf Consequences of the mapping in 1D} (a) Schematic representation of the motion of a
hole (circle) introduced in 1D AF (antiparallel arrows on nearest neighbors): although the first hop of the hole creates a spinon [wavy line, can further move (unshown) via 
spin exchange $\propto J$], all consecutive hops do not produce any extra 
spinons and the hole freely propagates as a `holon' $\propto t$  giving rise to spin-charge separation in 1D.
(b) Schematic representation of the motion of an orbital excitation (bold oval with the arrow in the upper bar) introduced in 1D AF:
as a result of the mapping, the case is qualitatively similar to (a) and the orbital excitation moves as an `orbiton',
giving rise to spin-orbital separation. 
[Figure and caption adopted from Fig. 3 of Ref.~\cite{Wohlfeld2011}]
}
\label{fig:3}
\end{figure}
\section{`Minimal' Spin-Orbital Model: Consequences of the Mapping}

Probably the most obvious advantage of the mapping lies in the reduction of the orbital problem to the one with fewer degrees of freedom. 
This, {\it inter alia}, means that the numerical solution of the problem can be performed on larger clusters.
However, a more important by-product of the mapping is related to the understanding of the orbital spectral function
-- for we can now refer to a large number of older studies discussing the problem of a single hole in the AF ground state.
For instance, restricting solely to the (most common in these model studies) hypercubic lattices, we infer that:

For the 1D case the $t$--$J$ model studies~\cite{Kim1996, Suzuura1997, Brunner2000, Kim2006, Moreno2013} have predicted the onset of the so-called spin-charge separation~\cite{Lieb1968, Luther1974}: the hole introduced into the the undoped AF first creates a single magnetic domain wall (spinon) but then freely moves as a `holon' in such a 1D AF.
While the hole carries both the spin and charge quantum number, the spinon (holon) solely carries the spin (charge) quantum number and 
one can talk of a spin-charge separation, cf.~Fig~\ref{fig:3}(a). For the orbital case this translates into the orbital excitation, which originally carries both
orbital and spin degree of freedom, splitting into the spinon and `pure' orbiton
collective excitations, cf.~Fig~\ref{fig:3}(b). We call this phenomenon a `spin-orbital separation'. 

\begin{figure}[t!]
\psfrag{J}{2J}
\includegraphics[width=1.0\columnwidth]{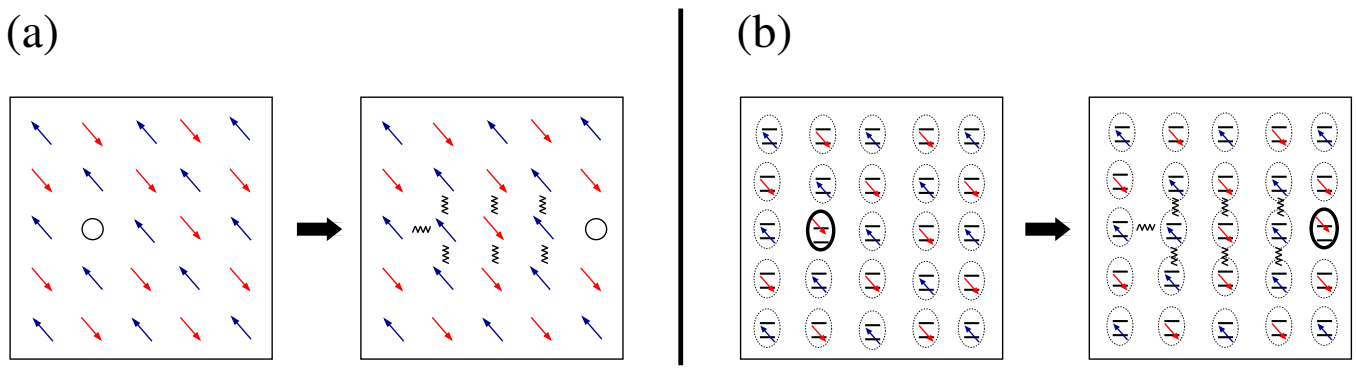}
\caption{{\bf Consequences of the mapping in 2D}
(a) Schematic representation of the motion of a hole (circle) introduced in 2D AF (antiparallel arrows on nearest neighbors):
the hole is subject to a string potential, for it creates spin defects (wavy lines) each time it makes a hop in a Neel-ordered 2D AF ground state
[it can become mobile solely due to the existence of the spin fluctuations which can heal these defects (unshown)].
(b) Schematic representation of the motion of an orbital excitation (bold oval with the arrow in the upper bar) introduced in 2D AF:
as a result of the mapping, the case is qualitatively similar to (a) and the orbital excitation is subject to a string potential in a 2D AF. 
[Figure and caption adopted from Fig. 2 of  Ref.~\cite{Wohlfeld2012}.]
}
\label{fig:4}
\end{figure}

On the other hand, a different situation occurs for the 2D or 3D lattice. In the simplest picture, the hole added to such an AF state 
introduces magnetic defects each time it makes a hop in a Neel-ordered 2D or 3D AF ground state~\cite{Brinkman1970}. As a result
the hole is confined in a string-potential, cf.~Fig~\ref{fig:4}(a). However, a more realistic picture reveals that these defects can be 
healed by quantum spin fluctuations $\propto J$ that the hole can become mobile -- albeit not on a scale
of the original free hopping $\propto t$ but rather $\propto J$~\cite{SchmittRink1988, Martinez1991}. In the literature the latter situation is often referred to as a `spin polaron'~\cite{Martinez1991}, since the hole becomes mobile by dressing with the collective magnetic excitations (magnons). Due to the mapping, a similar conclusion can be formulated
for the orbiton: the orbiton in a 2D / 3D AF is subject to a string potential [cf.~Fig~\ref{fig:4}(b)] and can become mobile solely by dressing with magnons.

\section{`Minimal' Spin-Orbital model:\\Compatibility of the Mapping Results with Other Approaches}


We mentioned above that the mapping is exact. Nevertheless, to fully understand the physics behind the mapping, 
it is necessary to compare the orbital spectral function calculated using the 
mapping against the exact numerical solutions as well as other available 
quasi-analytical solutions. {\it Below we concentrate solely on 1D, since such a comparison is basically only feasible in 1D}:
the numerically exact result experience large finite size effects, while the combination of cluster perturbation theory and exact diagonalization, 
that can be well-implemented in 1D (see below), fails in 2D~\cite{Kung2017}.
Moreover, there is also a well-known lack of reliable quasi-analytical approaches above 1D: (i)
Bethe Ansatz is developed solely for the 1D spin-orbital 
model~\cite{Sutherland1975, Zhang1998, Yamashita2000, Yu2000}, (ii) the large-$N$ mean-field approach~\cite{Baskaran1987, Affleck1988, Arovas1988} cannot be used for symmetry broken states 
that can be stabilized in two or higher dimensions, and (iii) the usual mean-field decoupling of spins and orbital~\cite{Oles2006} followed by a linear spin and / or 
orbital wave approach already does not lead to reasonable results in 1D.

\begin{figure}[t!]
\centering
\subfigure{
\includegraphics[width=0.5\linewidth]{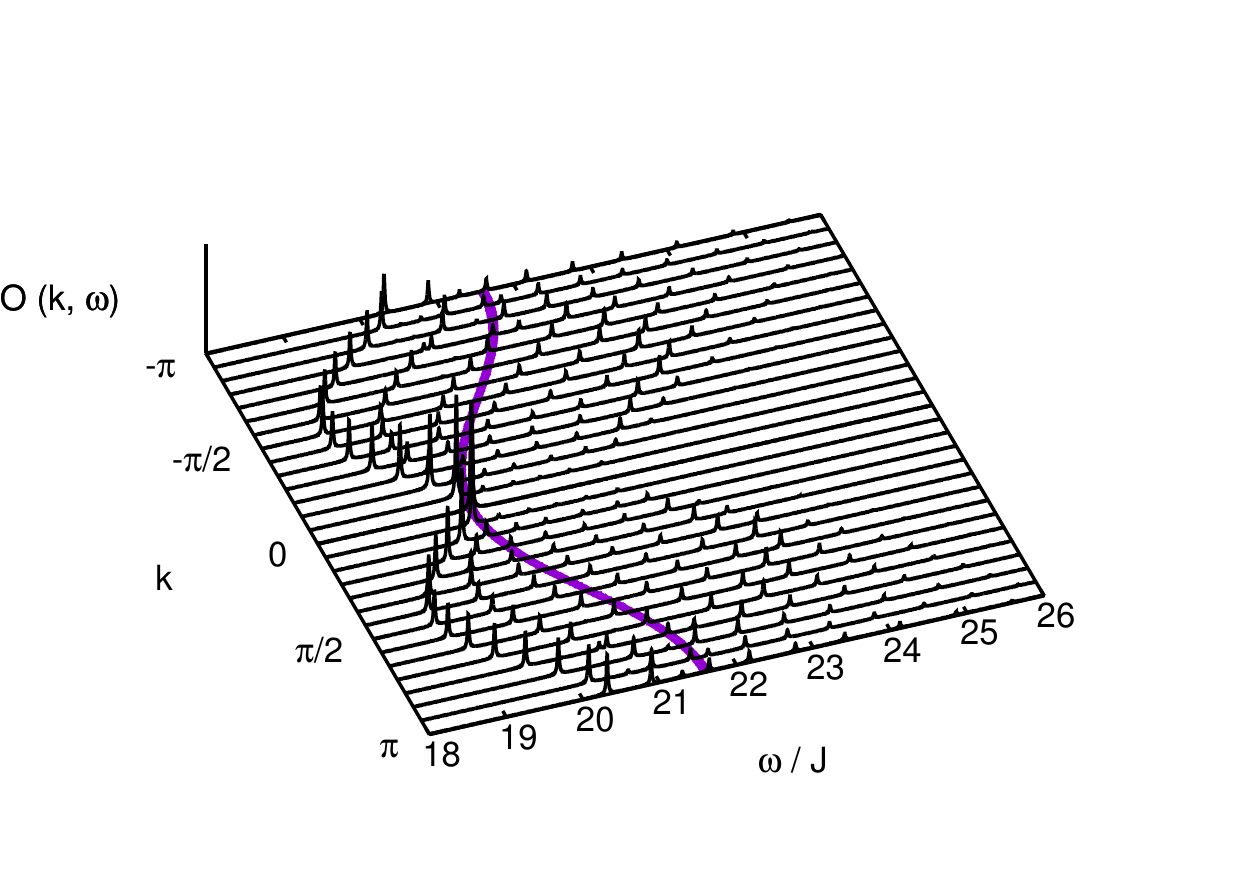}\llap{
  \parbox[b]{1.9in}{\normalsize{(T)}\\\rule{0ex}{2.2in}
  }}\label{graphJT}
}
\vskip0.4cm
\subfigure{
\includegraphics[width=0.4\linewidth]{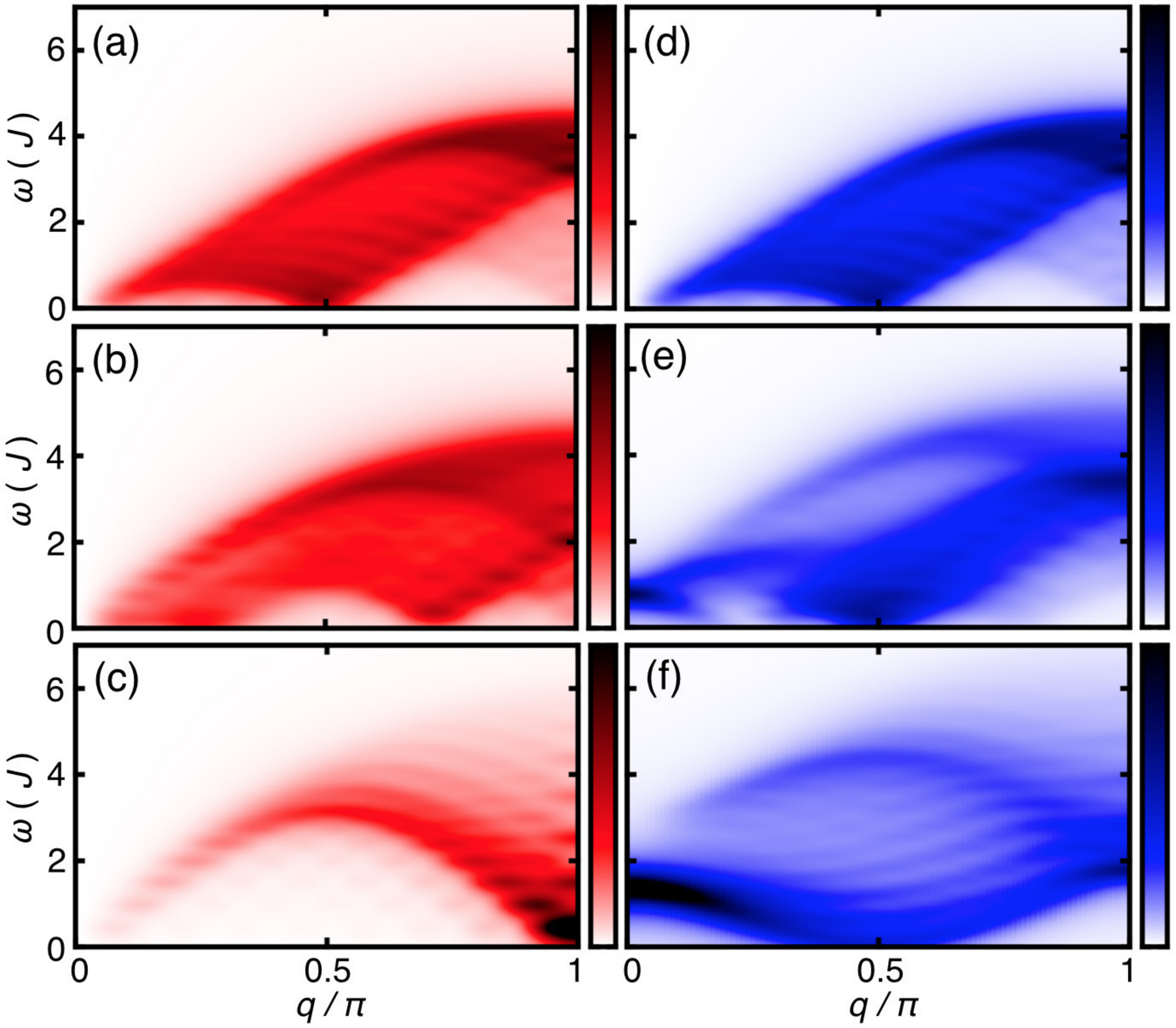}\llap{
  \parbox[b]{1.5in}{\normalsize{(L)}\\\rule{0ex}{2.8in}
  }}\label{graphJTfree}
}
\quad
\subfigure{
\includegraphics[width=0.4\linewidth]{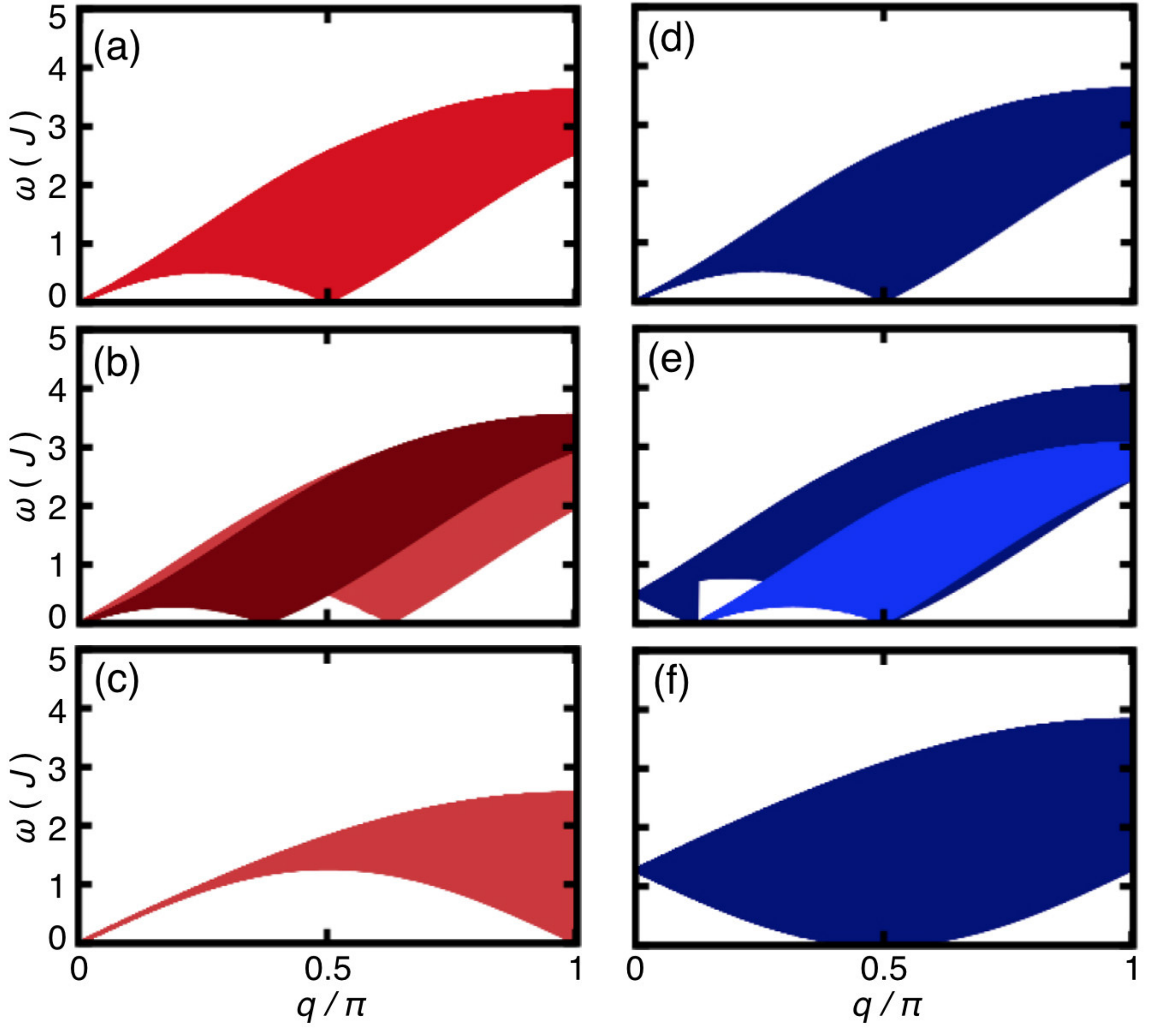}\llap{
  \parbox[b]{1.5in}{\normalsize{(R)}\\\rule{0ex}{2.8in}
  }}\label{graphSE}
}
\caption{{\bf Compatibility of the mapping results with other approaches}
(T) Comparison of the result from the mapping with the one obtained using the mean-field decoupling of spins and orbitals: orbital dynamical structure factor,  Eq.~(\ref{eq:spectral}), obtained via the 
mapping onto the $t$--$J$ model, Eq. (\ref{eq:spectraltJ}), evaluated 
using Lanczos exact diagonalization on a 28 site chain; a broadening 
$\eta = 0.06J$ and $E_z = 20J$; the thick
line shows result obtained in a mean-field and linear orbital wave approach, see text. 
(L) Result from the exact diagonalization on a 16 site chain supplemented by the cluster perturbation theory:
spin [(a)-(c)] and orbital [(d)-(f)] dynamical structure factor defined as in Eq. (\ref{eq:spectralS}) and  (\ref{eq:spectral}), respectively, at various $E_z$;
top panels: $E_z=0$, i.e. no orbital polarization;
middle panels: $E_z \approx 0.83 J $ leading to half polarized orbitals;
bottom panels: $E_z \approx 1.38 J$ leading to fully polarized orbitals;
see \cite{Chen2015} for more details.
(R) Result from the mean-field large-$N$ theory of constrained fermions:
compact support for spin [(a)-(c)] and orbital [(d)-(f)] spectra obtained by mean-field large-$N$ theory of constrained fermions at various $E_z$;
top panels: $E_z=0$, i.e. no orbital polarization;
middle panels: $E_z \approx 0.7 \frac{4 J}{\pi}$ leading to half polarized orbitals;
bottom panels: $E_z= \frac{4 J}{\pi}$ leading to fully polarized orbitals;
see \cite{Chen2015} for more details. [Panel (T) and caption adopted from Fig. 2 of Ref.~\cite{Wohlfeld2011}; note a different definition of the 
spin exchange $J$ used in the review and in Ref.~\cite{Wohlfeld2011} resulting in different energy scales in these manuscripts;
panels (L) and (R) adopted from Fig. 4 and 5 of Ref.~\cite{Chen2015}; $q$ on the OX axis of panels (L) and (R) denotes the momentum of the excitation (that is defined as $k$ in the main text)].
}
\label{fig:5}
\end{figure}

We start by showing the orbital spectral function calculated using the mapping for the case with $E_z = 20 J$ (i.e. the requirement $E_z \gg J$ is easily fulfilled), see
Fig.~\ref{fig:5}(T). The spectrum, calculated using Lanczos exact diagonalization method on a  28-site chain, 
is not formed by one single branch but instead it consists of multiple peaks (expected to merge into incoherent spectrum in the
thermodynamic limit) with a dominant feature at the lower edge of the spectrum. Using spin-charge separation Ansatz~\cite{Suzuura1997, Brunner2000}, adopted to the unusual case when the spin exchange is larger than the hopping, we observe that the bottom of the orbital spectral function is given by the dispersion relation 
$\omega\approx E_z-2t \sin |k|$ and is if a pure orbiton character. On the other hand the continuum
above that edge reflects both spinon and orbiton excitations -- except for the intermediate feature still well-visible within the continuum with 
the dispersion of the purely orbiton-character scaling as $\omega\approx  E_z+2t \sin |k|$.

We are now ready to compare the result obtained from the mapping with those that could be obtained using other approaches:

First, we compare the spectrum obtained using the mapping to the numerically exact solution of 
the original spin-orbital problem, i.e. Eqs.~(\ref{eq:h}-\ref{eq:spectral}).  
It occurs that both calculations, obtained using the Lanczos exact diagonalization suplemented by cluster perturbation theory~\cite{Maska1998, Senechal2000} on a 16-site chain,
give completely {\it identical} spectra, cf. Fig.~5(L)(f) which qualitatively shows the same spectrum as Fig.~5(T) (for a detailed comparison showing quantitative agreement cf.
Fig. 6(a) in \cite{Chen2015}). This {\it a posteriori} justifies the fact that the mapping is exact. 

Second, we discuss the orbital (and spin) spectral function (\ref{eq:spectral}) calculated
using a mean-field decoupling between spin and orbital operators in model Eq. (\ref{eq:h}), cf.~Ref.~\cite{Oles2006}.
This method is relatively straightforward, since, as a result of the Holstein-Primakoff transformation and linear orbital wave approximation that is employed to the decoupled orbital part of the spin-orbital Hamiltonian, it leads to a quadratic Hamiltonian which can be solved analytically. 
Such a decoupling was shown in the past to be relatively successful in correctly reproducing the ground and excited states 
of several spin-orbital models -- e.g. an analogous procedure was also performed for the 
magnons with FM and AO order in LaMnO$_3$ and KCuF$_3$ and gave a good 
agreement with experiment~\cite{Moussa1996, Lake2000}. Applying this procedure to the Hamiltonian (\ref{eq:h}) in the limit of $E_z \gg J$ 
leads to the orbital spectral function consisting 
of a single mode with a dispersion relation $\omega_{OW}(k)=E_z - \frac{1}{2}z J_{\rm OW}(1-\gamma_k)$.
Here $z$ is the coordination number, $\gamma_k$ is the lattice structure factor, and the 
effective orbital exchange constant $J_{\rm OW} = 4J \langle \psi_S | 
{\bf S}_{i } \cdot {\bf S}_{j} + \frac{1}{4} | \psi_S \rangle.$
Thus, in contrast to the exact case, the orbital excitation on the mean-field level is a quasiparticle with a
cosine-like dispersion with period $2 \pi$, cf.~the solid violet line in Fig.~\ref{fig:5}(T). 
This means that such a mean-field decoupling fails completely in the present case.

Third, we may wish to compare the results from the mapping with those based on the analytically exact solution following the Bethe Ansatz~\cite{Sutherland1975, Zhang1999, Yamashita2000}.
The latter predicts that a spin and orbital dynamical structure factor of model Eq.~(\ref{eq:h}) should be built out of three `flavorons' -- the collective
excitations of this model, each carrying both spin and orbital quantum number but (naturally) no charge. 
While this solution is exact and valid for {\it any} value of the crystal field $E_z$ in model Eq.~(\ref{eq:h}), not only understanding the nature of `flavorons' is a complex task but also 
it is extremely difficult to obtain the spectral information from any type of the Bethe Ansatz solution~\cite{Klauser2011}.
Therefore, we turn our attention to {\it another} mean-field decoupling, that is based on the so-called mean-field large-$N$ theory of constrained 
fermions~\cite{Baskaran1987, Affleck1988, Arovas1988}, can be used here and gives reasonable results. 
The method is based on firstly expressing the spin and orbital operators in Eq. (\ref{eq:h}) in terms of the so-called constrained 
fermions $f^\dag_{i \alpha \sigma}$, each carrying both orbital $\alpha=a, b$ and spin $\sigma$ degree of freedom 
and subject to the constraint $\sum_{ \alpha \sigma} {f}^\dagger_{i, \alpha \sigma} {f}_{i, \alpha \sigma} = 1$,
and then performing a mean-field decoupling of the obtained in this way interacting Hamiltonian.  
The latter decoupling is done in such a way that it preserves the $SU(4)$ symmetry of the interactions between spins and orbitals (see \cite{Chen2015} for more details)
and therefore it {\it may} give qualitatively correct results for systems without the spontaneously broken $SU(4)$ symmetry in their ground states~\cite{Baskaran1987, Affleck1988, Arovas1988}.
The resulting mean-field Hamiltonian $\mathcal{H}_{\rm MF}$ reads
\begin{equation}\label{eq:hfermion}
\mathcal{H}_{\rm MF} = \sum_{k , \sigma} \left(  \varepsilon_{k a } f^\dag_{k a \sigma}  f_{k a \sigma} +  \varepsilon_{k b} f^\dag_{k b \sigma}  f_{k b \sigma} \right),
\end{equation}  
where $ \varepsilon_{k a  / b} = -  4 \sqrt{2} J \cos (\delta_k)  \cos (k) / \pi \mp E_z /2$,
with $ \delta_k = \arcsin [ E_z \pi /  (4 J)] / 2$ when $E_z <  4 J / \pi$, and $\delta_k = \pi / 4$ when $E_z \ge 4 J / \pi$.
The crucial point here is that this mean-field Hamiltonian represents two doubly degenerate fermionic bands with their energies
$ \varepsilon_{k a  / b}$ separated by $E_z$. 
Due to the constraint of one fermion per site, the bands are filled up to the respective Fermi momenta:
$\pm k_F \mp \delta_k $ and $ \pm k_F \pm \delta_k$,
where $k_F =\pi/4$ is the Fermi momentum at $E_z=0$. Note that for $E_z=0$ the fermionic bands are four times degenerate and quarter-filled,
for $0<E_z <  4 J / \pi$ such orbital degeneracy is gone but all bands are occupied,
for $E_z <  4 J / \pi$ two of the four fermionic bands are fully occupied and two are completely empty. 
This result qualitatively agrees with the one obtained using the Lanczos exact diagonalization of the full Hamiltonian Eq. (\ref{eq:h})
which shows that for $E_z =0 $ the ground state shows AF and AO correlations without any orbital polarization,
for $0<E_z \lesssim 1.38 J $ the ground state is AF and orbitals are partially polarized,
and for $1.38 J  \lesssim E_z$ the ground state is AF and orbitals are fully polarized (FO order).

The main result here is that not only the ground state but also the low energy excited states can be qualitatively well-captured
by the large-$N$ mean-field Hamiltonian, Eq.~(\ref{eq:hfermion}). Moreover, this can be done for any value of the crystal field $E_z$.
Figure~\ref{fig:5}(L) shows the spin dynamical structure factor defined as 
\begin{align}
\label{eq:spectralS}
S(k, \omega)=\frac{1}{\pi} \lim_{\eta \rightarrow 0} 
\Im \left\langle \psi \left| S^-_{k} 
\frac{1}{\omega + E_{\psi}  - \mathcal{{H}} -
i \eta } S^+_k \right| \psi \right\rangle,
\end{align}
and the orbital dynamical structure factor, $O(k, \omega)$ [see Eq.~(\ref{eq:spectral})], calculated using the Lanczos exact diagonalization and the cluster perturbation theory~\cite{Maska1998, Senechal2000} of the full Hamiltonian Eq. (\ref{eq:h}) for three distinct values of the crystal field $E_z$ which correspond to three different regimes of the orbital polarization in the ground state.
This result is compared to the compact supports\footnote{As the mean-field approach cannot account for the spectral intensity~\cite{Raczkowski2013} we only concentrate
on the compact support.} of the spin and orbital spectra calculated using the mean-field Hamiltonian Eq. (\ref{eq:hfermion}),
see Fig.~\ref{fig:5}(R). The mean-field results agree well with the numerical simulations, for they can reproduce 
the shift in momentum of the zero-energy modes with the crystal field  $E_z$ as well as the overall bandwidths for the spin and orbital dynamical structure factors.
On a qualitative level, the only discrepancy between numerics and the mean-field approach is related to the missing low-intensity branch in the mean-field spectra:
that originates from the so-called $\omega$ flavoron~\cite{Zhang1999} in the Bethe Ansatz solution of the $SU(4)$ symmetric model 
and would involve four constrained fermions, which cannot be captured in the mean-field theory.

According to the large-$N$ mean-field approach, the excited states of the spin-orbital model Eq. (\ref{eq:h})
can be described in terms of noninteracting quasiparticles -- the constrained fermions $f^\dag_{k \alpha \sigma}$ where $k$ here is the (crystal) momentum. 
Since each of such momentum eigenstates carries both spin and orbital quantum numbers, therefore the spin
and orbital quantum number is {\it entangled} on each site, in a somewhat similar way as for the spin-orbital wave discussed in~Ref.~\cite{WenLong2012}.
While such a result agrees with the general notion on the nature of collective excitations based on the Bethe Ansatz~\cite{Sutherland1975, Zhang1999, Yamashita2000} and the numerical studies~\cite{Yu2000, Chen2007, Lundgren2012, WenLong2012} mostly done for the $E_z=0$ case,
it stays in contrast with the result from the mapping onto the $t$--$J$ model that suggests the spin-orbital {\it separation} when $E_z$ is large enough to fully polarize the orbital ground state.
How to reconcile that discrepancy? It turns out that the origin of this inconsistency lies in different definitions of the spin quantum number in these two approaches. 
Whereas in the large-$N$ mean-field approach electrons in both lower and upper orbitals carry spin and orbital quantum numbers, 
in the studies involving the mapping onto an effective $t$--$J$ model electrons in the lower (upper) orbital carry solely spin (orbital) quantum numbers, respectively. 
We note that the latter peculiar choice of the spin and orbital basis is possible only when the mapping of the spin-orbital model (having four degrees of freedom per site) to the $t$--$J$ model (having three degrees of freedom per site) is allowed in a ground state with fully polarized orbitals.

Irrespectively of the value of the crystal field $E_z$, in the mean-field picture the spin and orbital spectra could be understood as originating in the `particle-hole' excitations of the 
noninteracting constrained fermions $f^\dag_{k \alpha \sigma}$ across the Fermi level. Thus, both spin and orbital excitations are always {\it fractional}, since the original single spin or orbital flip (as measured in the spin or orbital structure factor) splits into the independent constrained fermions, each carrying both a fraction of the quantum number of an electron and a fraction of the quantum number associated with a single spin or orbital flip. For instance, an orbital-flip excitation in $O(k, \omega)$, which means changing the orbital quantum number by $\Delta T^z = 1$, 
fractionalizes into two fermions: one carrying $(S^z = \pm 1/2, T^z=1/2)$ quantum numbers, the other one carrying $(S^z =\pm 1/2, T^z=-1/2)$ quantum numbers, but (unlike electrons) both being chargeless.  

Interestingly, the latter situation stays in contrast with the one encountered for the single spin chain in the magnetic field, see \cite{Wohlfeld2015}. In principle, also for this case one could employ the large-$N$ mean-field theory. Unfortunately, this approach only works for spin chains in weak and moderate magnetic field. In this case the spin dynamical structure factor
calculated numerically and using this mean-field approach qualitatively agree. Nevertheless, the quantitative agreement between these two approaches is worse
than in the spin-orbital case. Moreover, the mean-field bands cannot accommodate more than  50\% polarized spins and the large-$N$ mean-field theory fails completely 
once the magnetic field is strong enough to polarize more than half of the spins in the ground state. 
Most importantly, however, the large-$N$ mean-field approach cannot explain the excitations above the critical magnetic field that fully polarizes the ground state and turns it into a ferromagnet
with no longer fractional excitations (the low lying excited states can be  well-described in terms $S=1$ quasiparticles -- magnons).
This means that the above mean-field theory works better for the spin-orbital model with the $SU(4)$ exchange interaction than for the $SU(2)$ spin chain and that
the spin chain in external magnetic field can, in this context, be regarded as more complex than a spin-orbital chain in a magnetic or crystal field. This is because the mean-field approximation for SU($N$) antiferromagnets gradually improves as $N$ becomes larger~\cite{Arovas1988, Auerbach1994} and becomes exact for $SU($N$)$ models when $N \rightarrow \infty$.

\section{Orbiton in a Quasi-1D Copper Oxide\\(and the Derivation of a Realistic Spin-Orbital Model)}


A crucial test for most of the physical theories is their experimental verification. In the case of the above mentioned theory of the propagation
of an orbiton in an antiferromagnet the situation is actually somewhat simpler. This is because `our' theory was actually constructed owing to the
access to the unpublished experimental results of J. Schlappa {\it et al.} in the end of 2009 and later published in Ref.~\cite{Schlappa2012}. It turned out that the experimental data that was presented at that time to us [see Fig~\ref{fig:6}(a) and discussion below], and which contained the spectrum of orbital excitations as observed with RIXS at Cu $L_3$ edge in a quasi-1D cuprate (Sr$_2$CuO$_3$),
had one very peculiar feature: large parts of the orbital spectrum seemed to be surprisingly similar to the photoemission spectrum of various quasi-1D cuprates~\cite{Kim1996, Kim2006}
that showed the onset of the spin-charge separation in 1D~\cite{Lieb1968, Luther1974}. It was this peculiar correspondence between these two seemingly different physical problems, that has inspired us in constructing the theory of the propagation of the orbiton in an antiferromagnet within the `minimal' spin-orbital model -- in particular the mapping between the spin-orbital and $t$--$J$ model problem. Therefore, it should not come as a surprise that the presented below theoretical model, that on a qualitative level should be regarded as an extension
to the `minimal' spin-orbital model, will be able to successfully explain the experimental data and confirm that the suggested spin-orbital separation can indeed be realised in a strongly correlated
crystal.

Let us now turn our attention to the understanding of that peculiar experimental RIXS data -- 
the part of the RIXS spectrum taken at Cu $L_3$ edge on Sr$_2$CuO$_3$ that is of interest to us is shown in Fig.~\ref{fig:6}(a). 
In the past this Mott insulating copper oxide had mostly been famous for an anomalously large spin exchange $J$, the extremely 1D character of its spin properties~\cite{Ami1995, Motoyama1996, Kojima1997}, and for the observation of spin-charge separation in the photoemission spectrum~\cite{Fujisawa1999}. It turns out that these are not the sole interesting features of this compound 
-- as Fig.~\ref{fig:6}(a) suggests there are three clear but peculiar characteristics of the RIXS spectrum of this compound:
(i) a weakly dispersive peak situated at around 1.85 eV at $k=0$, (ii) a strongly dispersive and pretty complex feature between ~2.1 and ~2.7 eV with the peak at $k=0$ at ca. 2.36 eV,
and (iii) a hardly dispersive peak at around 2.98 eV. How to interpret this spectrum? 

First of all, according to the theory of the RIXS cross section~\cite{Ishihara2000, Veenendaal2006, Forte2008, Ament2011, Marra2012} and in the widely-used fast collision approximation, one should think of RIXS at Cu $L_3$ edge as being directly sensitive to the orbital dynamical structure factor $O(k, \omega)$ in this energy range. Since in the simplest ionic picture the ground state orbital of the Cu$^{2+}$ ion located in the tetragonal ($D_{4h}$) symmetry of the CuO$_3$ chains realized in Sr$_2$CuO$_3$ is the $3d_{x^2-y^2} \equiv x^2-y^2$ orbital that is occupied by a single hole\footnote{In what follows the hole language is used.}, there should be four distinct orbital excitations (also called $dd$ excitations) accessible within the $3d$ shell (due to the tetragonal crystal field the cubic harmonics basis is used): 
$3d_{xz} \equiv xz$,  $3d_{yz} \equiv yz $, $3d_{xy} \equiv xy$ and $3d_{3z^2-r^2} \equiv 3z^2-r^2$. As a result of that we obtain the following equation for the RIXS cross section at Cu $L_3$ edge
in the interesting energy range:
\begin{align}\label{eq:final}
I ({ k},\omega)  &= |L_{xy} (k) |^2 O_{xy} (k, \omega) + |L_{xz} (k) |^2 O_{xz} (k, \omega) +  |L_{yz} (k) |^2 O_{yz} (k, \omega) 
+ |L_{3z^2-r^2} (k) |^2  O_{3z^2-r^2} (k, \omega),
\end{align}
where $O_{\alpha} (k, \omega)$ is the orbital dynamical structure factor related to a single excitation to one of the $3d$ orbitals ($\alpha=xz, yz, xy, 3z^2-r^2$),
and $L_{\alpha} (k)$ is the local RIXS form factor which follows from the dipole and fast collision approximation to the RIXS process, see Ref.~\cite{Wohlfeld2013} for details. The dependence of these factors
on the crystal momentum, explained in detail in Ref.~\cite{Wohlfeld2013}, follows from the fact that a change in the transferred momentum in the RIXS process leads to a change in the photon polarization vectors -- and that influences the cross section, cf.~Refs.~\cite{Veenendaal2006, Ament2011, Marra2012}. 

A simple calculation, which takes into account not only the crystal field but also the covalency effects due to the hybridization of the electrons on the copper ion with the four 
nearest neighbor oxygen ions, suggests that the ${xy}$ (${3z^2-r^2}$) is an excited orbital with lowest (highest) energy and that the energies of the ${xz}$ and ${yz}$ 
orbitals lie in between the other two orbitals and are degenerate. Such a structure was basically confirmed by the {\it ab-initio} quantum chemistry calculations. The obtained
energies of the orbital excitations, $E_{xy}=1.62$ eV, $E_{xz/yz}=2.25/2.32$ eV, $E_{3z^2-r^2}= 2.66$ eV match, within a ~15\% error bar, the position of the three most
intensive peaks found in the experimental spectrum at $k=0$, cf.~Fig.~\ref{fig:6}(a). Moreover, when the local RIXS form factor is taken into account, 
then the highest energy part of the RIXS spectrum can be surprisingly well reproduced using the `local picture', i.e. assuming that the orbital
excitations is dispersionless and that $  O_{3z^2-r^2} (k, \omega) = \delta(\omega - E_{3z^2-r^2})$, cf. the highest energy part of the experimental and theoretical spectrum
presented in Fig.~\ref{fig:6}(a) and (c). In fact, we have checked that, according to the DFT calculations, the hybridization of the $3z^2-r^2$ orbital
with the nearest neigbor oxygen orbitals is anomalously small in Sr$_2$CuO$_3$, i.e. it is much smaller than the simple Slater-Koster scheme would suggest~\cite{Slater1954}.
This further confirms that the momentum-dependence of the intensity of the $3z^2-r^2$ peak is solely due to the local RIXS form factor.

The main problem here concerns the calculation of the orbital dynamical structure factor for the $xy$ and $xz / yz$ orbital excitations, i.e. obtaining a model
which could explain the onset of the weak dispersion of the ${xy}$ orbital and the much stronger and more complex dispersion associated with 
the ${xz} / {yz}$ orbital. Let us first speculate on the possible origin of the onset of the latter dispersion.
We observe that lower branch seems to have a dominant $\pi$ period component and the upper branch seems to have a rather particular shape which altogether
leads to the onset of two `oval'-like features in the RIXS spectrum between ~2.1 and ~2.7 eV. On a qualitative level this is very similar to the theoretical
orbital spectrum calculated for the `minimal' spin-orbital model, see Fig.~\ref{fig:5}(T). Encouraged by this positive results we derived and solved in Ref.~\cite{Wohlfeld2013} a realistic spin-orbital model 
that could describe the propagation of both the ${xy}$ and the ${xz} / {yz}$ orbital excitation and is able to give a good quantitative agreement with the experiment. 
Here we summarize the crucial steps needed to obtain the orbital spectrum of that realistic spin-orbital model:
\begin{itemize}
\item Our starting point is the proper {\it charge transfer model} which, as already pointed long time ago by Zaanen, Sawatzky, and Allen~\cite{Zaanen1985}, should correctly describe
the low energy physics of the copper oxides. The crucial feature of this class of models is best understood in the hole language: it corresponds
to the fact that in the copper oxides it costs less energy to put an extra hole on the unfilled oxygen $2p$ shell (charge transfer $\Delta$) than 
on the single-occupied copper ${x^2-y^2}$ orbital (Hubbard $U$). Therefore, the oxygen degrees of freedom cannot be as easily integrated out as in the case
of the `standard' correlated systems with the low energy physics happenning only on the transition metal ions. The basic form of this 1D model for the CuO$_3$ chains in Sr$_2$CuO$_3$, including the suggested values of its parameters, was put forward by R. Neudert {\it et al.}~\cite{Neudert2000}. In Refs.~\cite{Schlappa2012} and \cite{Wohlfeld2013} we extended 
this model, in order to include the copper ${xy}$ and the ${xz}$ as well as those oxygen $2p$ orbitals which, according to the Slater-Koster scheme~\cite{Slater1954}, should strongly hybridize with these two $3d$ orbitals. Thus, in total, the model contains three $3d$ orbitals per copper ion and three $2p$ orbitals per oxygen ion ($2p_x \equiv p_x$, $2p_y \equiv p_y$, and $2p_z \equiv p_z$), see Eqs. (2-5) in~\cite{Wohlfeld2013}, and 18 model parameters. The values of the model parameters mostly follow Ref.~\cite{Neudert2000} with a few of them taken from the in-house LDA calculations or estimated
following Ref.~\cite{Grant1992}. We note that, as we choose that the  CuO$_3$ chains lie along the $x$ direction, we can safely assume that the ${yz}$ orbital should be nondispersive, since 
the hopping from the ${yz}$ orbital along the chain direction should be strongly suppressed~\cite{Slater1954}. Hence, we can write that
$O_{yz} (k, \omega)  = \delta(\omega - E_{yz}) $. 
\item As both $U$ and $\Delta$ are much larger than all the hopping elements ($t_n$) of the Hamiltonian we can hugely {\it simplify the charge transfer
model by considering perturbation theory in $t_n / \Delta$ and $t_n/U$}. 
\item In the zeroth order in the perturbation theory in hoppings $t_n$ of the charge transfer model and in the regime
of one hole per copper site, the large Coulomb repulsion $U$ and the charge transfer energy $\Delta$ cause Sr$_2$CuO$_3$ to be a Mott insulator.
This is because, in the zeroth order approximation in the perturbation theory in hopping $t_n$ and in the regime
of one hole per copper site, there is one hole localized in the $x^2-y^2$ orbital at each copper site $i$.
\item The second order perturbation theory\footnote{Contributions from first and other odd orders of perturbation theory vanish for this model.} 
in the hoppings $t_n$ leads to a relatively strong renormalization of the on-site energies of the orbital excitations due to the formation
of the bonding and antibonding states between the copper orbital and its nearest neighbor oxygen orbitals. 
However, since in the final calculations the on-site energies of the $xz$ and $xy$ orbital excitations will anyway be taken from the {\it ab-initio} quantum chemistry
calculations (see below), we can safely disregard this contribution.
\item The most interesting contribution arises from the fourth order in the perturbation theory in hoppings $t_n$ -- these are the so-called superexchange
processes in transition metal oxides~\cite{Zaanen1988}. This contribution defines the bulk part of the spin-orbital model and can be split into
three distinct parts:
\begin{itemize}
\item When only the ground state  ${x^2-y^2}$ orbital is occupied by a single hole on two nearest neighbor copper sites -- this leads to the following {\it spin} superexchange
Hamiltonian:
\begin{align}
\label{eq:h1}
\mathcal{{H}}_{\rm spin}&=J (1+R) \sum_{\langle i, j \rangle } \mathcal{P}_{i, j} \left({\bf S}_{i } \cdot {\bf S}_{j} -
\frac{1}{4} \right), 
\end{align}
where ${\bf S}_{i }$ is the $S=1/2$ spin operator for spins on the copper site $i$, 
the projection operator $ \mathcal{P}_{i, j}$ makes sure that there are no orbital excitations along the $\langle i, j \rangle$ bond,
the spin exchange is defined as $J=\frac{4{t}^4_{\sigma}}{(\Delta_x+V_{dp})^2} \frac{1}{U}$,
and $R=\frac{2U}{2\Delta_x+U_p}$ is the factor responsible for the onset of the superexchange processes on the oxygen ion.
Here ${t}_{\sigma}$ is the (renormalized due to the covalency effects, see Ref.~\cite{Wohlfeld2013} for more details) 
hopping element between the ${x^2-y^2}$ orbital and the nearest neighbor $p_x$ orbital in the CuO$_3$ chain,
$\Delta_x$ is the charge transfer energy for putting a hole into the $p_x$ orbital, $V_{dp}$ the Coulomb repulsion between holes on the nearest neigbor copper
and oxygen ions (as already defined $U$ is the Hubbard repulsion on the copper ion).
\begin{figure}[t!]
\centering
\psfrag{J}{2J}
\includegraphics[width=0.33\columnwidth]{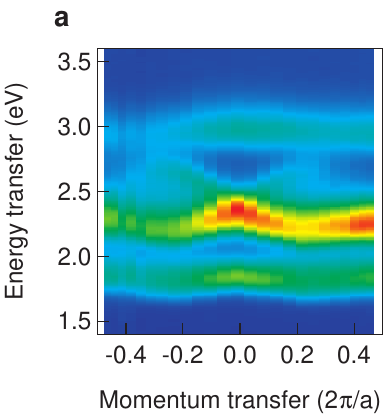} 
\includegraphics[width=0.265\columnwidth]{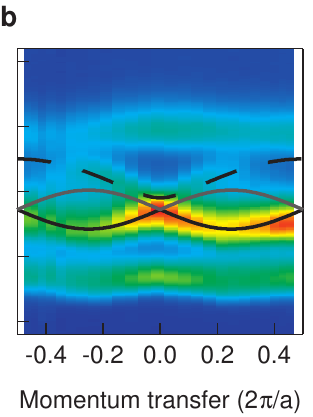} 
\includegraphics[width=0.387\columnwidth]{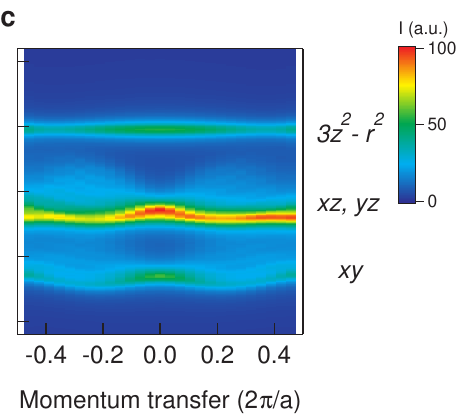} 
\caption{{\bf Orbiton in a quasi-1D copper oxide in experiment and theory} (a) Experimental spectrum containing four orbital excitations [$3d_{xy}$, $3d_{xz}$, $3d_{yz}$, $3d_{3z^2-r^2}$, see legend on the right hand side of panel (c)] as obtained using resonant inelastic x-ray scattering at Cu $L_3$ edge of Sr$_2$CuO$_3$.
Note that not all momenta (see white space in the spectrum close to $k \sim \pm \pi$)
in the first Brillouin zone of Sr$_2$CuO$_3$ are reached for the photons with energy $\sim930$ eV at Cu $L_3$ edge RIXS~\cite{Ament2011}.
(b) Same as (a) but with the dispersion relations given by the spin-orbital separation Ansatz showing the dispersion of the pure orbiton (solid line) and the edge
of the spinon-orbiton (dashed line) continuum of the $3d_{xz}$ orbital excitation (see text for more details). 
(c) Theoretical spectrum of orbital excitations as obtained using the spin-orbital model developed for Sr$_2$CuO$_3$ (see text for further details).
[Panels and caption adopted from Fig.~4 of Ref.~\cite{Schlappa2012}.]
}
\label{fig:6}
\end{figure}
The value of the spin exchange calculated in this way  ($J (1+R) \approx 0.24 $eV) 
well agrees with the one obtained from inelastic neutron scattering experiment $J=0.24$ eV~\cite{Walters2009}
as well as with the presented here resonant inelastic x-ray scattering at Cu $L_3$ edge which gives $J \approx 0.25$ eV (unshown low energy part of the RIXS spectrum, see Fig.~3 of \cite{Schlappa2012}).

\item When one copper site has a hole in the copper ${xz}$ orbital and its nearest neighbor copper site has the hole in the ground state ${x^2-y^2}$ orbital we obtain
the following spin-orbital Hamiltonian:
\begin{align}\label{eq:xz}
\mathcal{H}_{xz}= \sum_{\langle {i}, {j} \rangle } \left({\bf S}_{ i} \cdot {\bf S}_{ j} + A \right) \left[ B {T}^z_{ i} {T}^z_{ j} +\frac{C}{2} \left( {T}^+_{ i} {T}^-_{ j} + {T}^-_{ i} {T}^+_{ j} \right) + D \right],
\end{align}
where ${\bf T}_{i }$ is the $T=1/2$ orbital peudospin operator
with ${T}^z_{ i}$ is equal to the difference between occupation on the ${x^2-y^2}$ and the ${xz}$ orbital [${T}^z_{ i}= (n_{i, {x^2-y^2}} - n_{i, {xz}}) /2$].
All parameters of the model are defined in detail in Ref.~\cite{Wohlfeld2013}. Here we only concentrate on parameter $C=J_{xz} (R_1 + R_2 + r_1+r_2)$, 
as this is the crucial parameter, being responsible for the hopping of the orbital excitations between nearest neigbor copper sites. 
It turns out that this model parameter depends on the orbital exchange $J_{xz} = \frac{\left(2{t}_{\pi}\bar{t}_{\sigma}\right)^2}{(\Delta_z+V_{dp})(\Delta_x+V_{dp})} \frac{1}{U}$,
and on the four factors ($R_1$, $R_2$, $r_1$, $r_2$) which arise due to four distinct superexchange processes -- two from the low spin intermediate
states with two holes on the oxygen or copper sites and two from the high spin intermediate states with two holes on the oxygen or copper sites (they depend on the Hund's exchange 
and on the charge transfer energy, see \cite{Wohlfeld2013} for more details).
Finally, ${t}_{\pi}$ is the hopping element between the ${xz}$ orbital and the nearest neighbor $p_z$ orbital in the CuO$_3$ chain,
$\Delta_z$ is the charge transfer energy for putting a hole into the $p_z$ orbital.
\item When one copper site has a hole in the copper ${xy}$ orbital and its nearest neighbor copper site has the hole in the ground state ${x^2-y^2}$ orbital 
we obtain the following spin-orbital Hamiltonian:
\begin{align}\label{eq:xy}
\mathcal{H}_{xy}= \sum_{\langle {i}, {j} \rangle } \left( {\bf S}_{ i} \cdot {\bf S}_{ j} + A' \right) \left[ B' {T'}^z_{ i} {T'}^z_{ j} +\frac{C'}{2} \left( {T'}^+_{ i} {T'}^-_{ j} + {T'}^-_{ i} {T'}^+_{ j} \right) + D' \right],
\end{align}
where ${\bf T'}_{i }$ is the $T'=1/2$ orbital peudospin operator
with ${T'}^z_{ i}$ is equal to the difference between occupation on the ${x^2-y^2}$ and the ${xy}$ orbital [${T'}^z_{ i}= (n_{i, {x^2-y^2}} - n_{i, {xy}}) /2$].
Again all parameters of the model are defined in detail in Ref.~\cite{Wohlfeld2013} and we only write down explicitly the parameter $C'=J_{xy} (R'_1 + R'_2 + r'_1+r'_2)$
with $J_{xy} = \frac{\left(2 \bar{t}_{\pi} \bar{t}_{\sigma}\right)^2}{(\Delta_y+V_{dp})(\Delta_x+V_{dp})} \frac{1}{U}$,
the four factors ($R'_1$, $R'_2$, $r'_1$, $r'_2$) again arising due to four distinct superexchange processes,
and $\bar{t}_{\pi}$ being the (renormalized due to the covalency effects, see Ref.~\cite{Wohlfeld2013} for more details) hopping element 
between the ${xy}$ orbital and the nearest neighbor $p_y$ orbital in the CuO$_3$ chain,
$\Delta_y$ -- the charge transfer energy for putting a hole into the $p_y$ orbital.
\end{itemize}
\item Having derived the relevant spin-orbital model, we are now ready to calculate the relevant orbital dynamical structure factor --
for the single orbital excitation of the ${xz}$ character
\begin{align} \label{eq:spectral_xz}
O_{xz}(k, \omega)=\frac{1}{\pi} \lim_{\eta \rightarrow 0} 
\Im \left\langle \psi \left| T^-_{k} 
\frac{1}{\omega + E_{\psi}  - \mathcal{{H}}_{\rm spin} -  \mathcal{{H}}_{xz} +
i \eta } T^+_k \right| \psi \right\rangle,
\end{align}
as well as for the single orbital excitation of the ${xy}$ character
\begin{align} \label{eq:spectral_xy}
O_{xy}(k, \omega)=\frac{1}{\pi} \lim_{\eta \rightarrow 0} 
\Im \left\langle \psi \left| T'^-_{k} 
\frac{1}{\omega + E_{\psi}  - \mathcal{{H}}_{\rm spin} -  \mathcal{{H}}_{xy} +
i \eta } T'^+_k \right| \psi \right\rangle,
\end{align}
where in both cases $|\psi \rangle$ and $ E_{\psi}$ is the ground state and energy without any orbital excitations present, i.e. it is the ground state of $\mathcal{{H}}_{\rm spin}$.
\item The above orbital structure factors are calculated using the mapping onto an effective $t$--$J$ model, separately for each case (i.e. ${xz}$ and ${xy}$ orbital
excitation).  While the mapping basically proceeds as introduced in Sec.~III, there are two somewhat important differences here. First, the mapping is not exact,
since in both cases the spin-orbital Hamiltonian allows for a propagation of the orbital excitation that does {\it not} conserve the spin quantum number of the electron
in the excited orbital. Fortunately, that process, which is due to finite Hund's exchange $J_H$, has a relatively small contribution (it scales as $\propto J_H / U \sim 0.1-0.2$)
and can be neglected. Second, also due to the finite Hund's exchange, the mapping leads to a small (~10\%) spin-dependence of the 
hoppings in the effective $t$--$J$ model -- which is also neglected. 

Consequently, the problem of the orbital dynamical structure factor for the single orbital excitation of the ${xz}$ character can be mapped 
onto the problem of a single hole in the effective $t$--$J$ model, Eqs. (\ref{eq:spectraltJ}-\ref{eq:tJ}) with its effective $t$ and $J$ parameters being replaced by:
$t \rightarrow (3R_1+R_2+3r_1+r_2) / 8 J_{xz}$ and $J \rightarrow J(1+R) / 2$. Similarly, the orbital dynamical structure factor for the ${xy}$
orbital excitations can be calculated using the effective $t$--$J$ model with $t \rightarrow (3R'_1+R'_2+3r'_1+r'_2) / 8 J_{xy}$ (and $J$ as for the ${xz}$ case).
Moreover, in the definition of the spectral function, Eq.~(\ref{eq:spectraltJ}), we replace the on-site cost of creating a single orbital excitation, which was equal to the crystal
field in the `minimal' model, by its value as obtained from the {\it ab-initio} quantum chemistry calculations: $E_z \rightarrow E_{xz}$  and $E_z \rightarrow E_{xy}$,
respectively for each case. 
\item The two problems of a single hole in the effective $t$--$J$ models are solved using Lanczos exact diagonalization on a finite cluster.
\end{itemize}

The theoretical RIXS cross section, which follows Eq.~(\ref{eq:final}) and {\it inter alia} includes the orbital dynamical structure factors for the $xz$ and $xy$ orbitals
as explained above, is shown in Fig.~\ref{fig:6}(c). We observe that not only qualitative but also to a large extent quantitative 
agreement with the experimental spectra presented in~Fig.~\ref{fig:6}(a). The sine-like shape of the experimentally observed 
dispersion relation of the $xz$ and the $xy$ orbitals, such as period $\pi$ and minima at $\pm \pi/2$, is reproduced by the theoretical calculations. 
The same can also be said about the  width and shape of the `oval'-like bands above the dominant $xz$ dispersive peak. 
Finally, also the relative intensities of the particular features of the theoretical and experimental spectra agree.
The main discrepancy between the experiment and theory is related to the smaller dispersion in the theoretical calculations
than in the experiment. This might for instance be due to the small but finite spin-orbit coupling
in the 3$d$ shell which would mix the $xz$ and $yz$ orbital excitations
and could lead to a finite dispersion in the $yz$ orbital channel. 

The strong covalency effects lead to the effective hopping element $t$ in the 1D $t$--$J$ model being much larger for the $xz$ orbiton 
than for the $xy$ orbiton. Consequently the $xz$ orbiton is more mobile than the $xy$ orbiton and the 
characteristic features of the spin-orbital separation spectrum, well-known from Secs.~III and IV, are much better visible for the 
$xz$ orbiton than for the $xy$ orbiton in the RIXS spectrum, cf.~Fig~\ref{fig:6}(a), (c) with Fig.~\ref{fig:5}(T).
This is even better visible, when the already used spin-charge separation Ansatz~\cite{Suzuura1997, Brunner2000}
is adopted to the effective $t$--$J$ model for the $xz$ orbtion (`spin-orbital separation Ansatz'), cf. Fig.~\ref{fig:6}(b). 
Here we observe that indeed the bottom of the experimental $xz$ orbital spectral function is given by the dispersion relation 
$E_{xz}-2t \sin |k|$ (where $t \rightarrow (3R_1+R_2+3r_1+r_2) / 8 J_{xz}$, see above) and is if a pure orbiton character, see the solid line in Fig.~\ref{fig:6}(b).
The higher energy part of the $xz$ spectral function is more complex: it contains another feature of the purely orbiton character with the dispersion given
by $E_{xz}+2t \sin |k|$ and an upper edge of the joint spinon-orbiton continuum that is given by $E_{xz}+ \sqrt{4J^2+4t^2 + 8tJ \cos k}$ (where $J \rightarrow J(1+R) / 2$, see above).
Altogether, this additional calculations further confirms that RIXS on Cu $L_3$ edge of Sr$_2$CuO$_3$ has observed the orbiton excitation.

Could other theoretical scenarios explain the experimental data equally well? 
Of course we could never exclude such a scenario entirely -- however, a good agreement between theoretical modelling and experiments makes the task of finding such alternative frameworks rather challenging. 
In Ref.~\cite{Wohlfeld2013} few of these are discussed: (i) the onset of the dispersion as a consequence of the momentum-dependence of the local RIXS form factors,
(ii) a linear orbital wave approximation of the above-derived spin-orbital model, 
(iii) an observation of the spin-charge separation using RIXS, or (iv) the onset of the large dispersion following the large hopping elements between 
the oxygen $2p$ orbitals. It is suggested in \cite{Wohlfeld2013} that all of these theories cannot explain the experimental data equally well as the presented above spin-orbital model with
the onset of the spin-orbital separation in this quasi-1D cuprate.  

\section{Orbiton in a Ladder-like Copper Oxide\\(and the Robustness of Spin-Orbital Separation)}


\begin{figure}[t!]
\centering
\subfigure{
\includegraphics[width=0.43\linewidth]{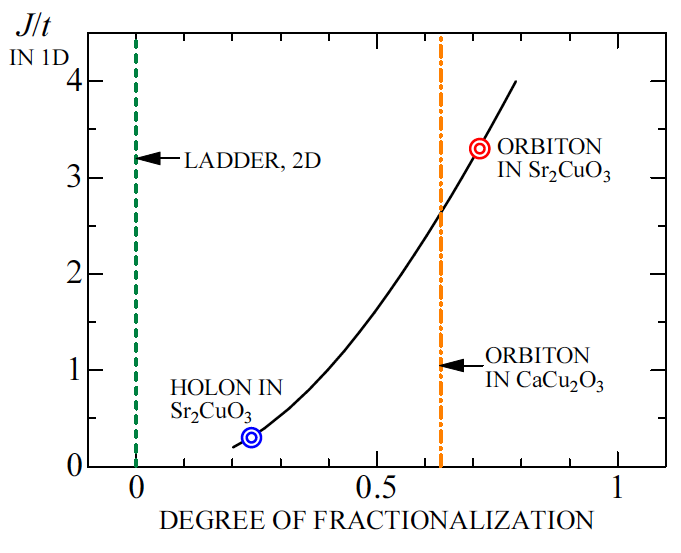}
  \label{graphJT}
}
\quad \quad \quad
\subfigure{
\includegraphics[width=0.43\linewidth]{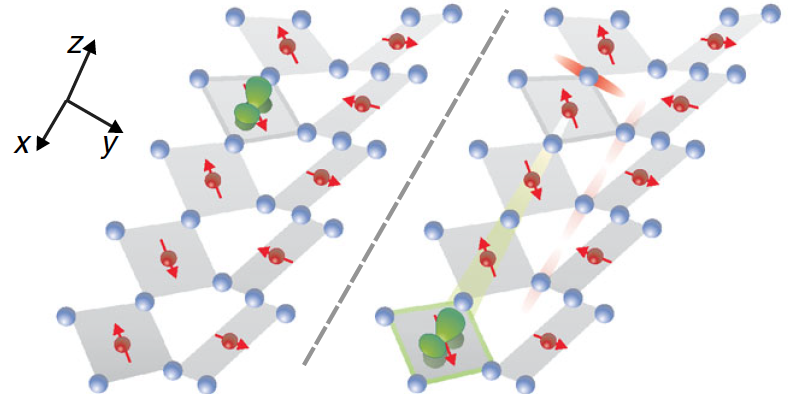}
  \label{graphJTfree}
}
\caption{{\bf Understanding spin-orbital separation in a ladder--like copper oxide} 
(${\rm a} \equiv {\rm left} $) Degree of fractionalization in various effective $t$--$J$ models 
as obtained using the finite size scaled exact diagonalization: for the
isotropic ladder $t$--$J$ model (dashed line), for the
anisotropic ladder $t$--$J$ describing the spin-orbital
separation in CaCu$_2$O$_3$ (dot dashed line), 
for the ideal 1D $t$--$J$ model calculated as a function of $J/t$ (solid line). 
$J/t \approx 0.4$ ($J/t \approx 3.3$) describes the spin-charge (spin-orbital) separation in
Sr$_2$CuO$_3$.
(${\rm b} \equiv {\rm right} $) Cartoon view of the orbiton propagation in the buckled two-leg ladder CaCu$_2$O$_3$:
when the ${xz}$ orbiton (marked in green) moves (compare left and right panels), it creates just one magnetic domain wall not only because interleg 
domain walls (light red) can be neglected (due to very weak spin interaction along the rung) but also because it can 
move {\it solely} along one of the legs of the ladders (due to the directional hopping of the ${xz}$ orbital).
[Panels and caption adopted from Fig.~4 and Fig.~1 of Ref.~\cite{Bisogni2015}]
}
\label{fig:7}
\end{figure}

The experimentally observed spectrum of orbital excitations in the quasi-1D copper oxide, Sr$_2$CuO$_3$, can be
very nicely understood using the 1D spin-orbital model. This {\it inter alia} lead to the first ever unambiguous observation of the orbiton
dispersion and of the spin-orbital separation. One can wonder whether this is the only compound for which this novel physics can be observed.
Below we give a very brief overview of the RIXS experiment on another copper oxide, CaCu$_2$O$_3$,
and show that also that RIXS spectrum shows signatures of the orbiton dispersion and, despite its `less-1D' structure, of the spin-orbital separation physics.

CaCu$_2$O$_3$ is a buckled two-leg spin ladder system that contains corner-sharing CuO$_4$ plaquettes~\cite{Lake2010, Bordas2005}.
This compound is very similar to Sr$_2$CuO$_3$: due to strong correlations it is a Mott insulator with the localized
holes on Cu$^{2+}$ ions in the $x^2-y^2$ orbitals. The superexchange processes are also allowed in this compound.
They lead to the strong, though around twice smaller than in Sr$_2$CuO$_3$, Heisenberg-like 
spin interaction between the $S=1/2$ spins of the localised holes along the leg direction of the ladder.  
Moreover, due to the buckling of the ladders leading to the onset of weak ferromagnetic exchange processes, 
the spin interaction across the rung of the ladder is relatively weak (~10 times weaker
than along the rung) which makes this ladder compound `more 1D' and closer to Sr$_2$CuO$_3$ than one 
would naively expect judging from the ladder-like coordination of the copper and oxygen atoms in its crystal structure~\cite{Bordas2005}.

The RIXS experiment, that was performed on the Cu $L_3$ edge of CaCu$_2$O$_3$, reveals a relatively similar spectrum
of the orbital excitations as in the case of Sr$_2$CuO$_3$, see Fig.~2 and~3 of Ref.~\cite{Bisogni2015}. Again there are at least three well-distinguishable peaks
between ca. 1.5 and 2.5 eV energy range. By comparing their energies to the quantum chemistry {\it ab-initio} calculations~\cite{Huang2011} 
one can easily assign the peaks lowest (highest) in energy to the $xy$ ($3z^2-r^2$) orbital excitation. These turn out to be nondispersive,
for a theoretical RIXS cross section which assumes purely local orbital excitations well describes that part of the experimental spectrum.
On the other hand, the part of the RIXS spectrum which is associated with the $xz / yz$ orbital excitation seems to show nonnegligible dispersion relation.

Can one reproduce the observed experimental behaviour, in particular the nondispersive character of the $xy$ orbital excitation and the dispersion of the $xz / yz$ orbital excitation,
with a theoretical model? It turns out that this indeed {\it is} possible (see excellent agreement between theory and experiment shown in Fig.~3 of Ref.~\cite{Bisogni2015}) 
and that one can do that following similar steps as proposed in Sec.~VI for Sr$_2$CuO$_3$.
In order not to dive too much into the details below we merely mention the main differences between these two cases:
\begin{itemize}
\item The charge transfer model is defined on a (buckled) two-leg ladder instead of the 1D chain. As obtaining the correct values of the hopping parameters in this buckled system
is a crucial step in modelling the orbital spectrum of this compound, the hopping parameters of its tight-binding part are obtained from the in-house DFT calculations
(FPLO code, cf.~Ref.~\cite{Koepernik1999}). We note in passing that the other parameters of the model follow (as for Sr$_2$CuO$_3$) Ref.~\cite{Neudert2000}, since
we do not expect the Hubbard $U$ or the charge transfer energies to substantially changed between different copper oxides.
\item Consequently, the spin model, obtained from charge transfer model in the fourth order perturbation theory, is defined on a two-leg ladder and reads:
\begin{align}
\label{eq:hSE_CCO}
\mathcal{{H}}_{\rm spin}&=J_{\rm rung} \sum_{\langle {\bf i}, {\bf j} \rangle  || {\rm rung}} \mathcal{P}_{{\bf i}, {\bf j}} \left({\bf S}_{\bf i } \cdot {\bf S}_{\bf j} -
\frac{1}{4} \right)+ J_{\rm leg} \sum_{\langle {\bf i}, {\bf j} \rangle  || {\rm leg}} \mathcal{P}_{{\bf i}, {\bf j}} \left({\bf S}_{\bf i } \cdot {\bf S}_{\bf j} -
\frac{1}{4} \right), 
\end{align}
where $\mathcal{P}_{{\bf i}, {\bf j}}$ is defined as in Eq.~(\ref{eq:h1}) and due to the buckling the model is very anisotropic, i.e. $J_{\rm rung} \approx J_{\rm leg} /12$, cf. Refs~\cite{Bordas2005, Lake2000}.
\item An important difference arises when deriving the spin-orbital model for the $xy$ orbital excitation.
It turns out that the hopping along the leg of the ladder is blocked due to a rather peculiar interplay of the Pauli
principle and a covalency effect following the strong interladder hopping between the $p_y$
oxygen orbital in the leg and the $x^2-y^2$ copper orbital in
the neighboring ladder. As explained in detail in \cite{Bisogni2015} the coherent
travel of the $xy$ orbital along the leg is suppressed and therefore we can safely model using it a purely local picture
(the motion along the rung is not resolved in the RIXS experiment, since there is no transferred momentum along the rung in the chosen
geometry of the RIXS experiment).
\item Although for different reasons, the situation for the $xz$ orbital excitation is somewhat opposite to the $xy$ case: the spin-orbital superexchange processes
are allowed along the leg (and no covalency and Pauli principle effects can hinder its motion along the leg), the hopping along the rung
vanishes completely due to the `one-dimensional' character of the hopping from the $xz$ orbital [cf. Fig.~\ref{fig:7}(b)].
As a result we obtain the purely 1D spin-orbital Hamiltonian for the $xz$ orbital orbital excitation:
\begin{align}\label{eq:xz_CCO}
\mathcal{H}_{xz}= \sum_{\langle {\bf i}, {\bf j} \rangle || {\rm leg} } \left({\bf S}_{\bf  i} \cdot {\bf S}_{\bf  j} + A \right) \left[ B {T}^z_{\bf  i} {T}^z_{\bf  j} +\frac{C}{2} \left( {T}^+_{\bf  i} {T}^-_{\bf  j} + 
{T}^-_{\bf  i} {T}^+_{\bf  j} \right) + D \right],
\end{align}
where (naturally) all the parameters of the model are differ quantitatively w.r.t. the Sr$_2$CuO$_3$ case.
\end{itemize}

The effective $t$--$J$ model, which follows through the mapping from the above model Eqs.~(\ref{eq:hSE_CCO}-\ref{eq:xz_CCO}) and describes the propagation of the $xz$ orbiton in CaCu$_2$O$_3$ is not strictly 1D: although the hopping of a hole has a strictly 1D character, there is a small but finite spin interaction 
along the ladder rungs, see Eq.~\ref{eq:hSE_CCO}. Nevertheless, the orbiton spectrum calculated from this model using exact diagonalization on a $14 \times 2$ ladder cluster seems to be very similar to the one of the completely 1D model and a spin-orbital separation Ansatz seems to well-describe the spectrum, cf.~Fig.~3 of \cite{Bisogni2015}. 
On the other hand, the analysis of the magnetic spectrum of CaCu$_2$O$_3$ suggests that for the energy scale $E \ll J_{rung}$ 
spinons are {\it not} the `correct' collective excitations~\cite{Lake2010}. 

In order to establish whether the spin-orbital separation really takes place in this model on the here relevant energy scale, 
we define the so-called `degree of fractionalization' as
the ratio between the correlation $\lambda (r)$ at $r = \infty$ and $r = 1$ with
$\lambda$ defined as the correlation between the orbital (or charge in the case of spin-charge separation) degree of freedom
at site $i = 1$ and the spin at site $i + r$.
This quantity expresses how well the spinon and the orbiton (or
holon) are separated from each other and it ranges from 0
(not separated) to 1 (fully separated).
Numerically the degree of fractionalization is calculated using the finite size scaling, which was needed in order to estimate the value
of the correlations at infinite distance. The results, obtained for various models and shown in Fig.~\ref{fig:7}(a), suggest that the degree of fractionalization
is finite for the model describing the $xz$ orbiton in CaCu$_2$O$_3$ and it is just a bit weaker than for the strictly 1D case of Sr$_2$CuO$_3$.

Interestingly, the degree of fractionalization is stronger in the case of the spin-orbital separation in a not-strictly 1D compound (CaCu$_2$O$_3$)  
than in the case of the spin-charge separation in completely 1D case (Sr$_2$CuO$_3$), cf. Fig.~\ref{fig:7}(a).
There are three reasons for that 
(i) the spin excitations in the buckled spin ladder CaCu$_2$O$_3$ are essentially spinons on the here relevant energy scale, cf. Fig.~\ref{fig:7}(b);
(ii) the motion of the orbital excitation can be 1D due to the typical
directional character of orbital hoppings, cf. Fig.~\ref{fig:7}(b);
(iii) the different ratio of the spinon and orbiton velocities versus spinon and holon velocities
in these two cases means that the spinon can move away much quicker from the
orbiton than from the holon, allowing for an `easier'
separation from the spinon in the spin-orbital case.
While the first condition requires 1D spin exchange
interactions and is similarly valid for the spin-charge
separation phenomenon, the other two conditions are
generic to the spin-orbital separation phenomenon.
This means that the spin-orbital separation is in general more robust than the spin-charge separation
and can be more easily observed in systems which are not strictly 1D.

\section{Orbiton in a Quasi-2D Iridium Oxide\\(and the Role of Jahn-Teller Effect)}


So far we have shown that the orbiton can indeed be observed in two copper oxides: one with a purely 1D (Sr$_2$CuO$_3$, see Sec.~VII)
and the other one with a two-leg ladder geometry (CaCu$_2$O$_3$, see Sec.~VIII). As the latter one had a relatively small coupling
between the legs of the ladders, effectively both cases were somewhat similar: as a result of the `mapping' between a spin-orbital
problem and an effective $t$--$J$-like problem the orbiton could be seen as a hole that could easily become 
mobile in an antiferromagnet -- due to the spin-orbital separation effect. 
Could the latter situation be observed in one of the quasi-2D transition metal oxides?
 
The theoretical discussion presented in Secs.~III-IV suggested that such a situation could be quite different in higher dimensions -- 
for instance already in 2D. Even though the mapping still holds in 2D, the orbiton should
become less mobile due to the lack of spin-orbital separation in 2D and the dressing of the mobile orbiton with spin fluctuations.
For example such a simple but relatively realistic 2D spin-orbital model with two inequivalent orbital hoppings was proposed in Ref.~\cite{Wohlfeld2012}.
Mapping that model onto the $t$--$J$ model lead to effective hopping of the orbiton $t=2t_1 t_2 / U$ in the effective $t$--$J$
model. Here $t_1$ ($t_2$) denote the bare hopping of an electron between the nearest neighbor ground (excited) orbitals and
$U$ is the onsite Hubbard repulsion. For quasi-2D copper oxides we would typically have for the ground state $x^2-y^2$ orbital and one of the
the excited $t_{2g}$ orbitals $t_2 \approx t_1/2 $. This gives $J \approx 4t$ in the effective $t$--$J$ model, since $J \approx 4t_1^2/U$. 
Solving such a 2D $t$--$J$ model, using the well-suited to the problem mapping onto the spin polaron problem and the self-consistent Born approximation~\cite{Martinez1991},
would give the strongly renormalized bandwidth of the orbiton dispersion  $W \approx 0.125J$, cf.~Fig. 1(c)-(d) of Ref.~\cite{Wohlfeld2012}.
Unfortunately, this would mean that for the quasi-2D copper oxides the orbiton bandwidth $W~10$ meV. It therefore
should not surprise us that so far the orbiton dispersion has not been detected in the quasi-2D copper oxides\footnote{A small dispersion of the orbital excitations was detected in the {\it doped}
quasi-2D copper oxides~\cite{Ellis2015}. However, due to doping, that dispersion might be closely related to the motion of doped holes in these compounds.}
and the $dd$ excitations in these systems are typically assumed to be completely local, cf.~Ref.~\cite{Moretti2011}. 
 
\begin{figure}[t!]
\centering
\includegraphics[width=0.62\columnwidth]{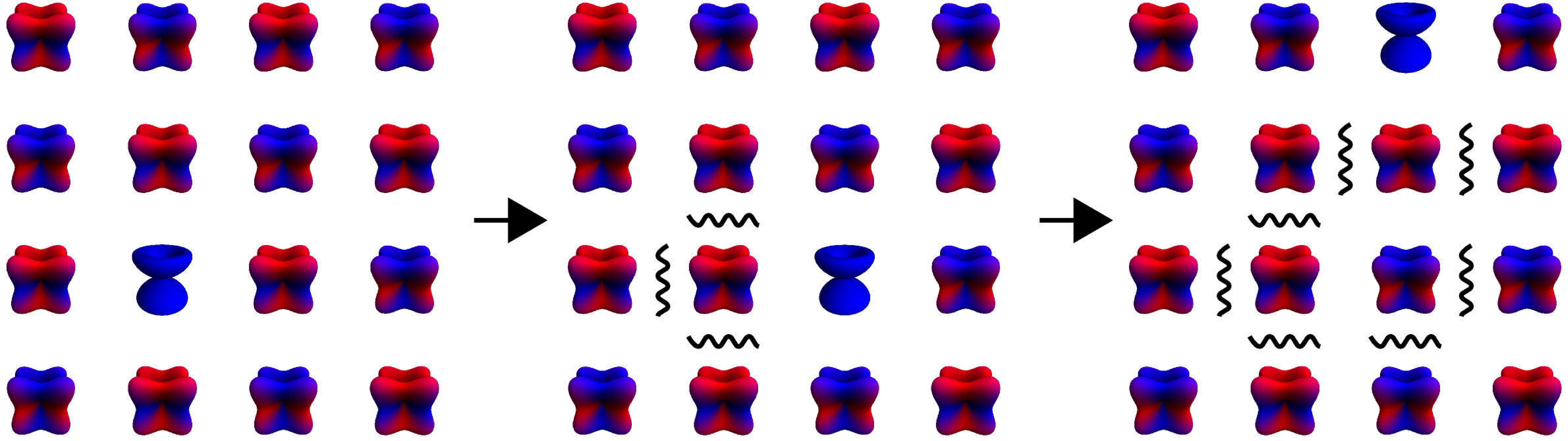}
  \label{cartoonHoppinga}
\\
\vskip 1cm
 \includegraphics[width=0.62\linewidth]{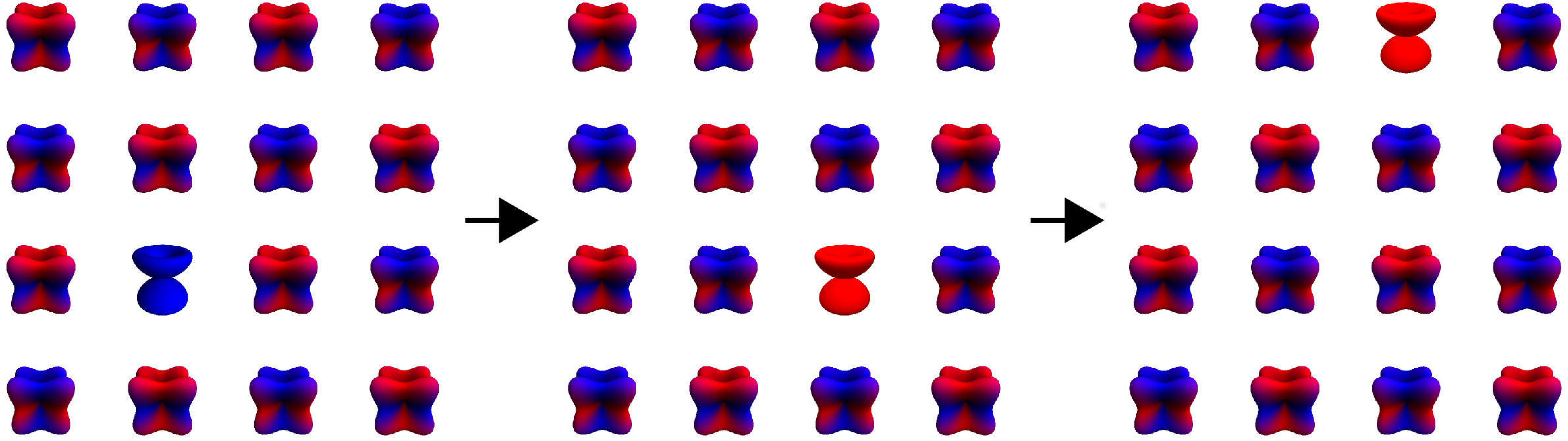}
    \label{cartoonHoppingb}
      \caption{{\bf Propagation of a $j=3/2$ orbiton in the $j=1/2$ antiferromagnetic order of Sr$_2$IrO$_4$}
(${\rm a} \equiv {\rm top} $) Cartoon showing such a propagation via the polaronic hopping (due to Jahn-Teller effect $or$ superexchange): a $j=3/2$ orbiton with the $j_z=-3/2$ quantum number (left panel) does not change its $j_z$ quantum number
during the hopping process to the nearest neighbor sites (middle / right panels) and thus the $j=1/2$ magnons are created at each step
of the hopping of an orbiton (wiggle lines on middle and right panels).
(${\rm b} \equiv {\rm bottom} $) Cartoon showing such a propagation via the free hopping (solely due to Jahn-Teller effect): a $j=3/2$ orbiton with the $j_z=-3/2$ quantum number (left panel) hops to the nearest neighbor site 
and acquires $j_z=3/2$ quantum number (middle panel). Note that in this case the $j=1/2$ magnons are {\it not} created in the system (middle / right panels).
[Panels and caption adopted from Fig.~4 of Ref.~\cite{Plotnikova2016}]
\label{fig:8}}
\end{figure}

It turns out that orbitons can be observed in a different class of quasi-2D transition metal oxides: the family of the quasi-2D `214' iridium oxides, best
exemplified by Sr$_2$IrO$_4$ or 
Ba$_2$IrO$_4$,\footnote{Often jointly called as the `2-1-4 iridates' (which differentiates them from the `2-1-3 iridates' with the honeycomb lattices).} 
which have a surprisingly similar crystal and electronic structure as 
the `famous' quasi-2D copper oxides, see Refs.~\cite{Kim2008, Wang2011}. 
Although in the ionic picture Ir$^{4+}$ leads to a $5d^5$ configuration, the tetragonal crystal field and the relatively strong spin-orbit coupling $\xi $ of the $5d$ iridium shell (ca. 0.4 eV, cf.~\cite{Kim2014}) hugely simplify the problem. In fact, in the hole picture the iridium ion has a single hole in the doubly degenerate 
ground state with an effective $j =1/2$ total angular momentum [$j=l+s$ where $l=1$ is the (effective) orbital angular momentum of the $t_{2g}$ subshell
and $s=1/2$ is the spin angular momentum of a single hole]. The strong correlations localize the holes in the iridium ions leading to a Mott insulating ground state and to
the low energy physics being governed by an effective interaction between $j=1/2$ spin-orbitals. The latter follows from the superexchange processes,
which are predominantly between nearest neighbor iridium ions and which, on a square lattice formed by the iridium ions in the IrO$_2$ planes of Sr$_2$IrO$_4$,
occur to be of the Heisenberg type~\cite{Jackeli2009}. Consequently the ground state is magnetic, it is an antiferromagnetic order formed by the $j=1/2$ spin-orbitals
(see Fig.~\ref{fig:8}).  The low energy excitations carry $j_z=1$ quantum number and are of collective type -- to show their similarity with those known for the 
2D $S=1/2$ Heisenberg antiferromagnet, they are called $j=1/2$ magnons. The $j=1/2$ magnon dispersion was detected by the recent RIXS experiment~\cite{Kim2012} 
and is well described using a linear spin wave approximation. The qualitative similarity of that low energy physics with the well-known La$_2$CuO$_4$ is striking.

The low energy excited states on the iridium ion, that lie above the manifold spanned by the states carrying the $j=1/2$ spin-orbital quantum number
(i.e. the AF ordered ground state and the magnons), resemble the orbital excitations of the cuprates.
Indeed, locally i.e. on a single iridium ion, there are four states carrying $j=3/2$ quantum number (i.e. having a different total angular momentum as the ground state) 
which are separated from the $j=1/2$ states by a gap $3 \xi /2$ due the spin-orbit coupling $\xi$. 
We can call these excited states `$j=3/2$ orbitals'.  Moreover, one expects that a hole in one of the $j=3/2$ excited orbitals can swap its places with the nearest neighbor hole in the $j=1/2$ spin-orbital via a $`t^2/U'$ process. Consequently, due to these spin-orbital exchange processes, the $j=3/2$ orbital excitations can 
acquire collective nature. While in the literature this $j=3/2$ collective excitation is often called an `exciton' [see Refs.~\cite{Kim2012, Kim2014, Plotnikova2016}], 
it can equally well be called a `$j=3/2$ orbiton'.  Finally, as such an exchange process happens in an antiferromagnet, the $j=3/2$ orbiton has to strongly couple to the $j=1/2$ magnons, just
as could be expected from the `mapping' between such a spin-orbital problem and an effective $t$--$J$ problem, cf.~Fig.~\ref{fig:8}(a). 

This interesting but pretty complex physics of the quasi-2D iridates would probably not be worth studying, if not for the fact that it can be very nicely seen
in the RIXS experiment. Indeed, the RIXS spectrum taken on the iridium $L_3$ edge of Sr$_2$IrO$_4$~\cite{Kim2012, Kim2014}, reveals a rather complicated spectrum of the $j=3/2$ orbital excitations, cf.~Fig.~3 of Ref.~\cite{Kim2014} but with a clear onset of the strong dispersion and a large incoherent continuum. Nevertheless, the spectrum is quite well-described by the spin-orbital model which contains the spin-orbital exchange processes and the mapping onto an effective $t$--$J$ like problem, as described above~\cite{Kim2012, Kim2014}. The only discrepancy lies in the unexplained feature with the minimum at the $\Gamma$ point that was observed in the normal incidence RIXS spectrum, cf. Fig. 3(a) and 4(b) in Ref.~\cite{Kim2014}. It was the main task of Ref.~\cite{Plotnikova2016} to explain this feature. Thus, below we briefly show how to derive and solve the model which includes the propagation of the $j=3/2$ orbiton due to the Jahn-Teller effect:
\begin{itemize}
\item Our starting point here is the Jahn-Teller effect for the $t_{2_g}$ electrons. In this case, as a result of the electron-phonon interaction, 
the orbital operators ${\bf l }$ couple to the tetragonal phonon modes $Q_2$ and $Q_3$ (the
$e_g$ modes) and to the trigonal phonon modes $Q_4$, $Q_5$, and $Q_6$ (the $t_{2g}$ modes). After integrating out
the phonons, the Jahn-Teller interaction is expressed solely in terms of the (effective) orbital angular momentum operator ${\bf l}$~\cite{Kugel1982}: 
\begin{align}
\label{HJT}
\mathcal{H}_{\rm JT}=&V \sum\limits_{\langle {\bf i}, {\bf j} \rangle}\left[\bigl(l_{\bf i}^z\bigr)^2-\frac{2}{3}\right]\left[\bigl(l_{\bf j}^z\bigr)^2-\frac{2}{3}\right] +V \sum\limits_{\langle {\bf i},{\bf j} \rangle} 
\left[\bigl(l_{\bf i}^x\bigr)^2-\bigl(l_{\bf i}^y\bigr)^2\right] 
\left[\bigl(l_{\bf j}^x\bigr)^2-\bigl(l_{\bf j}^y\bigr)^2\right] + \kappa V \sum \limits_{\langle {\bf i},{\bf j} \rangle}\Big[\left(l_{\bf i}^xl_{\bf i}^y+l_{\bf i}^yl_{\bf i}^x\right)\times \nonumber \\
&(l_{\bf j}^xl_{\bf j}^y+l_{\bf j}^yl_{\bf j}^x)+...\Big]
\end{align}
where $V$ ($\kappa V$) is the strength of the Jahn-Teller interaction due to the coupling to the $t_{2g}$ ($e_g$) phonons.
Since the Jahn-Teller coupling constant for $t_{2g}$ phonons is typically much smaller one can safely set
$\kappa=0.1$. On the other hand, the precise value of the Jahn-Teller interaction is at present unclear and is left here as a free parameter.  
\item Due to the strong spin-orbit coupling, it is convenient to express 
the Jahn-Teller interaction $\mathcal{H}_{\rm JT}$ in the eigenbasis of the total angular momentum operator ${\bf j}={\bf l}+ {\bf s}$, see Ref.~\cite{Plotnikova2016}. The terms of the resulting
Hamiltonian can be grouped into three classes
\begin{align}
\mathcal{H}_{\rm JT}\! =\! \mathcal{H}_{\rm JT} (1/2, 1/2)\! +\! \mathcal{H}_{\rm JT}(3/2, 1/2)\!+\! \mathcal{H}_{\rm JT}(3/2, 3/2),
\end{align}
where the first terms $\mathcal{H}_{\rm JT}(1/2, 1/2)$ denote the
Jahn-Teller interaction between two $j=1/2$ total angular momenta, 
the terms $\mathcal{H}_{\rm  JT}(3/2, 1/2)$ describe the interaction between one $j=1/2$ and one
$j=3/2$ total angular momentum, and the last terms  $\mathcal{H}_{\rm JT}(3/2, 3/2)$ express the interactions
between two $j=3/2$ total angular momenta. Interestingly, the first terms vanish here, 
reflecting the well-known~\cite{Kugel1982} quenching of orbital physics within the $j=1/2$ subshell.
At the same time the last term $\mathcal{H}_{\rm JT}(3/2, 3/2)$
only contributes if a large number of $j=3/2$ states are present and is strongly suppressed at large $\xi$.
Consequently, these are the middle terms, $\mathcal{H}_{\rm  JT}(3/2, 1/2)$, that become 
relevant when RIXS raises a single hole from the $j=1/2$ into a $j=3/2$ state~\cite{Kim2012, Kim2014}. 
\item Ultimately, as in the other section, we intend to calculate the spectral function for a single $j=3/2$ orbiton ($|{\chi}_{ {\bf k}, {j_z}} \rangle$, where $j_z = \pm 3/2, \pm 1/2$)
added to the AF $j=1/2$ ground state ($| 0 \rangle$) and with the dynamics given by the Jahn-Teller Hamiltonian  $\mathcal{H}_{\rm  JT}(3/2, 1/2)$
as well as the superexchange Hamiltonian known from Refs.~\cite{Kim2012, Kim2014} ($\mathcal{H}_{\rm SE}$),
\begin{flalign}
\label{green}
O({\bf k},\omega)=  \frac{1}{\pi} \lim_{\eta \rightarrow 0} 
\Im {\rm Tr}_{j_z} {\Big\langle 0 \Big| {\chi}_{ {\bf k}, j_z}\frac{1}{\omega-{\mathcal{H}_{\rm  JT}(3/2, 1/2)} - \mathcal{H}_{\rm SE} +i\delta}
{\chi}_{ {\bf k}, j_z}^\dagger \Big| 0 \Big\rangle}.
\end{flalign}
\item To this end, we map the above problem of the orbiton propagation 
directly onto an effective spin polaron problem~\cite{Martinez1991}.
We note that, although this mapping is inspired and is closely related to the mapping onto the $t$--$J$ model presented in Sec.~III,
it is also quite distinct: (i) on one hand, it is only valid in the presence of long range order;
(ii) on the other hand, it can allow for the change of the $j_z$ quantum number of the orbiton. 
As a result the Jahn-Teller Hamiltonian reads
\begin{align}
	    \label{H_finalmag}
	\mathcal{H}_{\rm  JT}(3/2, 1/2)&=
        \sum\limits_{{\bf k}}{ \hat{E}^{\rm JT}_{{\bf k}} 
          \hat{\chi}^\dagger_{\bf k} \hat{\chi}^{\phantom{\dagger}}_{{\bf k}} }	+ \sum\limits_{{\bf k}, {\bf q}}{\left(\hat{M}^{\rm JT}_{{\bf k},{\bf q}}
	\hat{\chi}^\dagger_{{\bf k}} \hat{\chi}^{\phantom{\dagger}}_{{\bf k}- {\bf q}}a^{\phantom{\dagger}}_{{\bf q}}+h.c.\right)},
\end{align}
where $ \hat{\chi}^\dagger_{\bf k}$ is a vector of four creation operators that create a $j=3/2$ orbiton with one of the four $j_z$ quantum numbers
and $a^{\phantom{\dagger}}_{{\bf q}}$ creates a magnon with momentum ${{\bf q}}$. The detailed expressions for the 
free hopping of the $j=3/2$ orbitons ($\hat{E}^{\rm JT}_{{\bf k}}$) as well as 
the interactions of the $j=3/2$ orbitons with magnons ($\hat{M}^{\rm JT}_{{\bf k},{\bf q}}$) are given in \cite{Plotnikova2016}.
There, also the detailed expression for $\mathcal{H}_{\rm SE}$ is given.
\item Finally, the orbital spectral function Eq. (\ref{green}), is calculated using, the well-suited to the polaronic problems, self-consistent Born approximation~\cite{Martinez1991}.
\end{itemize}

The orbiton spectral function calculated using not only the superexchange but also the Jahn-Teller model reproduces well the experimental RIXS spectrum of Refs.~\cite{Kim2012, Kim2014}, see Fig.~2 of \cite{Plotnikova2016}. In particular, due to the Jahn-Teller interaction an additional feature with the minimum at the $\Gamma$ point occurs in the spectrum
and therefore the experimental RIXS spectrum is fully explained. 

It is interesting to understand whether the Jahn-Teller effect can lead to a qualitatively distinct propagation of the $j=3/2$ orbiton than the superexchange mechanism.
It turns out that the difference between these two
mechanisms for orbiton propagation in the quasi-2D iridate is of fundamental nature. The crucial aspect
concerns the nearest-neighbor processes, which are depicted for superexchange and Jahn-Teller effect in Fig.~\ref{fig:8}. 
In superexchange, the $j=3/2$ orbiton propagates by exchanging place with an
$j=1/2$ spin-orbital while both conserve their $j_z$ quantum number. In the ground state given by the $j=1/2$ antiferromagnetic order, where nearest
neighbors are always of opposite $j_z$, this necessarily
creates or removes `defects', see Fig.~\ref{fig:8}(a), i.e. the $j=1/2$ magnons. 
On the other hand, the Jahn-Teller effect allows the $j=3/2$ orbiton
and the $j=1/2$ spin-orbital to flip their $j_z$ quantum numbers while exchanging places and this allows for the nearest
neighbor hopping of a $j=3/2$ orbiton without creating $j=1/2$ magnons, i.e. a free orbiton
dispersion, see Fig.~\ref{fig:8}(b). The origin of the difference lies in the fact that the electronic hopping
driving the superexchange conserves the $j_z$ quantum number, while the
lattice-mediated Jahn-Teller effect is insensitive to the orbital
phase. This allows $j_z$ to change during the Jahn-Teller--driven
propagation.

\section{Conclusions and New Perspectives for the Orbital Physics}

This review discusses the problem of the propagation of collective orbital excitation (orbiton)
in several distinct strongly correlated systems that, however, all bear an antiferromagnetic ground state.
From the theoretical point of view, {\it the most interesting result concerns the mapping
of the problem of an orbiton propagation in such an antiferromagnetic ground state 
onto a problem of a single hole in an effective half-filed $t$--$J$ model}~\cite{Wohlfeld2011}. 
That result has inspired few theoretical works on the problem of the so-called spin-orbital separation~\cite{Kumar2012, Brzezicki2014, Vieira2014} (as well as \cite{Chen2015,Wohlfeld2015})
and has greatly helped in understanding several experimental results on the orbiton dispersion found by the resonant inelastic x-ray scattering (RIXS)
experiments. The models describing these experiments on an almost quantitative level were discussed in detail in this review: 
in the quasi-1D cuprate~\cite{Schlappa2012, Wohlfeld2013}, in the ladder-like cuprate \cite{Bisogni2015}, and in the quasi-2D iridates~\cite{Kim2012, Kim2014} (as well as \cite{Plotnikova2016}). 
The  latter studies have also allowed us to learn more about the various aspects related to the orbiton propagation --  such as e.g.:
the importance of various superexchange processes for the realistic description of the orbiton motion \cite{Schlappa2012, Wohlfeld2013},
the robustness of the spin-orbital separation \cite{Bisogni2015}, or the role that could be played by the Jahn-Teller effect in the propagation of orbitons \cite{Plotnikova2016}.

What seem to be the most challenging problems in the coming years? 
Probably the most apparent problem, which directly follows from the studies presented in this review, concerns 
the {\it observation of the orbiton propagation in the quasi-2D cuprates}. As already mentioned, so far all of the orbital excitations 
observed in the RIXS experiments on these compounds seemed to have a purely local character. There is a general belief, that is also supported
by simple calculations~\cite{Wohlfeld2012}, that this is solely due to the too small resolution of the current RIXS experiments.
However, the newly installed RIXS ID-32 beamline in European Synchrotron Radiation Facility (ESRF), that had already been exploited to beautifully uncover the dispersion of magnons
in several cuprates~\cite{Peng2017}, should be capable in measuring the orbiton dispersion 
with a far better resolution.

Another open question is more theoretical---it concerns a far {\it better understanding the role played by the electron-phonon coupling in the orbiton propagation}. 
For instance, having seen how important the Jahn-Teller effect is for the orbiton propagation in the quasi-2D iridates, one
can wonder why the electron-phonon interaction was {\it not} included in the reported above
studies of the orbiton propagation in the quasi-1D or ladder-like cuprates. Moreover, although that was not discussed in the Introduction, 
it is well-known that the Jahn-Teller effect is extremely important for the understanding of the onset of orbitally ordered phases in several `classical' orbital systems, 
such as KCuF$_3$ or LaMnO$_3$~\cite{Kugel1982, Oles2005}. 

While a definite answer to this problem is not clear, one can speculate as follows:
The Jahn-Teller effect turned out to be important for the quasi-2D iridates, since it allowed
a channel of the `free' orbiton propagation between nearest neighbor sites. 
Note that, without Jahn-Teller and restricting solely to nearest neighbor processes, the orbiton 
is strongly coupled to magnons and in 2D can only move by dressing with magnons.
At the same time, in the case of the lower-dimensional 
cuprates, the orbiton is in any case relatively `free'---which is due to the spin-orbital separation.
Therefore, any extra channel that allows for `free' propagation is not that relevant.
Nevertheless, a thorough study on the role of the electron-phonon coupling on the orbiton propagation
is needed. Such a study should not only be contrasted with the findings presented in this review but also
with the earlier studies~\cite{Allen1999, vandenBrink2001, Schmidt2007} 
that suggested that the electron-phonon played 
an important (and destructive) role in the orbiton propagation~\cite{Allen1999, vandenBrink2001, Schmidt2007}. 

\section{Acknowledgments}

I would like to thank all of my collaborators who have greatly contributed to the understanding of the 
orbiton motion in the antiferromagnets---in alphabetical order these are: Valentina Bisogni, Jeroen van den Brink, Cheng-Chien Chen, 
Maria Daghofer, Tom Devereaux, Jochen Geck, Maurits Haverkort, Liviu Hozoi, Giniyat Khaliullin, Claude Monney, Satoshi Nishimoto, Ekaterina Paerschke (Plotnikova), 
Henrik Ronnow, Justina Schlappa, Thorsten Schmitt, Michel van Veenenedaal, Kejin Zhou (and many others). I thank Andrzej M. Ole\'s, as well as Tadeusz Doma\'nski, Tomasz Kostyrko, and Romuald Lema\'nski for the critical reading of my habilitation thesis---on which this manuscript is based. Finally, I am also grateful to Lucio Braicovich, Giacomo Ghiringhelli, and Markus Grueninger 
for several insightful discussions on the orbiton motion observed by RIXS. This work was supported by Narodowe Centrum Nauki (NCN, Poland) under Projects No.~2016/22/E/ST3/00560 and 2016/23/B/ST3/00839.


\end{document}